\documentclass [a4paper,11pt]{article}
\usepackage[english]{babel}
\usepackage{graphicx,amsmath,amsfonts,amssymb,slashed,upgreek,color,caption,amscd,cite,cancel}
\usepackage{cite}
\usepackage{framed}
\usepackage{multirow}
\usepackage{graphicx,subfigure}
\usepackage{multicol}

\newcommand{\cb}{\mathcal{C}}
\newcommand{\wb}{w_{\rm eff}}

\newcommand{\vv}{\vskip 2mm}

\def\geff{G_{\rm eff}}
\def\gn{G_{ N}}

\def\gs{\eta}
\def\mp{M_{\rm Pl}}

\newcommand{\fs}{f\sigma_8}

\newcommand{\be}{\begin{equation}}
\newcommand{\ee}{\end{equation}}
\newcommand{\beq}{\begin{equation}}
\newcommand{\eeq}{\end{equation}}
\newcommand{\bea}{\begin{eqnarray}}
\newcommand{\eea}{\end{eqnarray}}

\newcommand{\G}{G}

\newcommand{\pz}{p^{(0)}}
\newcommand{\pu}{p^{(1)}}
\newcommand{\pd}{p^{(2)}}

\newcommand{\R}{{}^{(3)}\!R}
\newcommand{\vn}{\vec \nabla}

\usepackage[stable]{footmisc}

\def\dkmu2{\delta K_{\mu \nu}\delta K^{\mu \nu}}
\def\pmu2{  \phi_{\mu \nu}\phi^{\mu \nu}}

\usepackage{a4wide}
\usepackage{fullpage}
\usepackage[utf8]{inputenc} 
\usepackage[english]{babel} 
\usepackage{color} 
\usepackage{epstopdf}
\usepackage{float} 
\usepackage[usenames,dvipsnames,svgnames,table]{xcolor}
\usepackage{multicol}
\usepackage{amsmath, amsfonts, amssymb, amsthm} 
\usepackage{bm} 
\usepackage{listings} 
\usepackage{mathrsfs} 
\usepackage{relsize} 
\usepackage{multirow}
\usepackage{pdfpages} 

\begin{document}

\begin{center}
\Large{\textbf{Phenomenology of dark energy:  \\ general features of large-scale perturbations}} \\[1cm]

\large{Louis P\`erenon$^{\rm a,b}$, Federico Piazza$^{\rm a,b}$, Christian Marinoni$^{\rm a,b}$ and Lam Hui$^{\rm c}$ }
\\[0.5cm]

\small{
\textit{$^{\rm a}$ Aix Marseille Universit\'e, CNRS, CPT, UMR 7332, 13288 Marseille,  France. \\
}

\textit{$^{\rm b}$Universit\'e de Toulon, CNRS, CPT, UMR 7332, 83957 La Garde,  France. \\
}

\textit{$^{\rm c}$ Physics Department and Institute for Strings, Cosmology, and Astroparticle Physics,\\
  Columbia University, New York, NY 10027, USA}
 }

\vspace{0.5cm}

\end{center}

\vspace{2cm}

\begin{abstract}
We present a systematic exploration of dark energy and modified
gravity models containing a single scalar field non-minimally coupled
to the metric. 
Even though the parameter space is large, by exploiting an effective
field theory (EFT) formulation 
and by  imposing simple physical constraints such as stability
conditions and (sub-)luminal propagation of perturbations, we arrive at a number of
generic predictions. 
(1) The linear growth rate of matter density fluctuations is generally suppressed compared to $\Lambda$CDM 
at intermediate redshifts ($0.5 \lesssim z \lesssim 1$), despite the introduction of an attractive
long-range scalar force. This is due to the fact that, in self-accelerating models,
the background gravitational coupling weakens at intermediate redshifts, over-compensating 
the effect of the  attractive scalar force. 
(2) At higher redshifts, the opposite happens; we identify a period of super-growth 
when the linear growth rate is larger than that predicted by $\Lambda$CDM.
(3) The gravitational slip parameter $\eta$---the ratio of the space part of
the metric perturbation to the time part---is bounded from above.
For Brans-Dicke-type theories $\eta$ is at most unity.
For more general theories, $\eta$ can exceed unity at intermediate redshifts, 
but not more than about $1.5$ if, at the same time, 
the linear growth rate is to be compatible with current observational
constraints. We caution against phenomenological parametrization of data
that do not correspond to predictions from viable physical theories.
We advocate the EFT approach as a way to constrain new physics
from future large-scale-structure data.
\end{abstract}

\newpage 
\tableofcontents

\vspace{.5cm}

\section{Introduction}\label{sec_1}

Understanding the nature of the present cosmic acceleration is an important and fascinating challenge. 
The standard paradigm---a cosmological constant $+$ Cold Dark Matter
within the framework of general relativity ($\Lambda$CDM)---has so far held up
remarkably well when tested against cosmological data~\cite{Betoule:2014frx,planck,VIPERS,Aubourg:2014yra}. 
This is especially true for data on the background expansion history.
Large scale structure data, in other words data that concern the
fluctuations, are improving in precision---current constraints
are broadly consistent with $\Lambda$CDM, although mild tensions exist
(see \emph{e.g.}~\cite{Samushia:2012iq,Steigerwald:2014ava, Ade:2015rim,Ade:2015fva}). New surveys, such as DES \cite{DES}, Euclid \cite{Euclid}, DESI \cite{DESI}, LSST \cite{LSST} and WFIRST \cite{WFIRST}, 
are expected to greatly tighten these constraints.
An important question is whether or how these observations can be used to
distinguish $\Lambda$CDM from other dark energy models~\cite{WHMH, guz, amen,  BBMV}.

From the theoretical perspective, any form of dark energy that is not the
cosmological constant would have fluctuations.
These dark energy fluctuations can couple to matter or not.
Or, in the frame where matter couples only to the metric (the frame we
use in this paper), these dark energy fluctuations can kinetically mix with
the metric fluctuations or not.
Models that have no such mixing are quintessence models---they
generally give predictions for structure growth fairly close to $\Lambda$CDM, especially if
the background expansion history is chosen to match to the observed
one, with the equation of state index close to $-1$.
Models that have such mixing are modified gravity models,
our primary interest in this paper.

Effective field theory provides a framework for systematically writing
down the action that governs the dynamics of such dark energy
fluctuations
\cite{Creminelli:2008wc,EFTOr,Bloomfield:2012ff,GLPV,Bloomfield:2013efa,PV,pheno,Gergely:2014rna,Gleyzes:2014rba,Gleyzes:2015pma} (see~\cite{eftcamb1,eftcamb2,eftcamb3} for a numerical implementation of this formalism).
The background cosmic expansion is treated as a given---matching that of the best-fit $\Lambda$CDM for instance---while the dark energy
fluctuations are encoded by a single scalar, the Goldstone boson
associated with spontaneously broken time-diffeomorphism in the gravity/dark energy sector.
In this paper, we are interested in the linear evolution of
fluctuations, thus we retain terms in the action up to quadratic
order in a gradient expansion. An important feature of the effective field theory for dark
energy is that the Ricci scalar comes with a general time-dependent
coupling. This allows the accelerating background expansion to
be driven by modified gravity effects, as opposed to what
resembles vacuum energy.

At the level of linear perturbations, non-standard gravitational
scenarios effectively result in a time- and scale-dependent modifications of the  Newton's constant $\geff$
and of the gravitational slip parameter $\gs$~\cite{Pogosian:2010tj}. 
The former quantity captures information about the way 
mass fluctuations  interact in the universe, while the gravitational slip parameter quantifies any 
nonstandard relation between the Newtonian potential $\Phi$ (time-time part of the
metric fluctuations)
and the curvature potential $\Psi$ (space-space part).

Models of dark energy contain in principle so many parameters that it might
seem hopeless to come up with robust, generic predictions for $G_{\rm eff}$,
$\gs$.
For example, \emph{Horndeski
  theory}~\cite{horndeski,Deffayet:2009mn}---the most general theory
containing one additional scalar degree of freedom $\phi$, with
second order equations of motion---depends on \emph{four} arbitrary functions of $\phi$ \emph{and} of the kinetic term $(\partial \phi)^2$. If we relax the condition of no-higher derivatives in the equations of motion to that of absence of pathological ghost instabilities, we end up with an even larger \emph{beyond Horndeski} set of theories, dependent on \emph{six} arbitrary functions of $\phi$ and $(\partial \phi)^2$~\cite{GLPV2,GLPV3}.

One of the main purposes of this paper is to show that, despite such
apparent freedom,  the behavior of $\geff(z)$ and $\gs(z)$ as
functions of the redshift\footnote{The scale dependence of $\geff$ and
  $\gs$ arises from (possibly time-dependent) mass terms for the dark
energy fluctuations. A natural value for the mass is the Hubble scale,
implying essentially no scale dependence in $\geff$ and $\gs$ for
fluctuations on scales much smaller than the Hubble radius.
An observable scale dependence for the growth of such fluctuations
can only arise if one introduces a mass scale higher than
Hubble. This is the case for chameleon models for instance \cite{WangHuiKhoury}.
We will not consider this possibility here.}
has definite features common to all healthy dark energy
models (models with no ghost and gradient instabilities, nor
superluminal propagation) within the vast Horndeski class. 
As a corollary, we show that popular, phenomenological
parameterizations of $\geff(z)$ and $\gs(z)$ do not capture
their redshift-dependence in actual, physical models.

The phenomenology of theories containing up to two derivatives in the
equations of motion has been explored in Ref.~\cite{pheno}, hereafter paper I, 
where a complete separation between background and perturbations
quantities has been obtained 
(see also~\cite{Sawicki}) and a systematic study of the growth index of matter density fluctuations has been initiated. 
As shown in paper I, the background evolution and the linear cosmological perturbations for any scalar-tensor theory of this type are entirely captured by one constant and five functions of time,\footnote{With respect to the equivalent ``$\alpha$-parameterization" introduced in Ref.~\cite{Sawicki} (see  App.~\ref{app-a}), the ``$\mu$-parameterization" used here is more theory-oriented: as summarized in the text, our couplings are in direct correspondence with the galileon and/or Horndeski Lagrangians of progressively higher order. }
\begin{equation} \label{couplings}
\left\{\Omega_m^{0},\ H(t),\ \mu(t),\ \mu_2(t), \ \mu_3(t),\ \epsilon_4(t) \right\}\, .
\end{equation}
The above ingredients allow one to span the entire set of Horndeski
theories, which can be seen as generalizations of
galileon-theories~\cite{NRT}. Since also models of massive
gravity~\cite{deRham:2010ik} and bi-gravity~\cite{Hassan:2011ea}
reduce to the galileons in the relevant (``decoupling")
limit~\cite{deRham:2010tw,Fasiello:2013woa}, 
aspects of these models are captured by our analysis.
In the above, $\Omega_m^{0} = \rho_m(t_0)/(3 mp^2 H_0^2)$ is the
present fractional energy density of non-relativistic matter; $H(t)$,
the Hubble parameter, encapsulates the background
expansion history;  the $\mu$s are parameters of the perturbation
sector with the dimension of mass, typically of order Hubble: $\mu(t)$
is the non-minimal coupling of Brans-Dicke (\emph{BD}) theories,
$\mu_3$ appears in cubic galileon- and Horndeski-3 (\emph{H3}) theories,
$\mu_2$ affects the sound speed of the scalar fluctuations, but has
otherwise no bearing on the linear growth of matter fluctuations;
$\epsilon_4$ is a dimensionless order one parameter
present in galileon-Horndeski 4 and 5 (\emph{H45}) Lagrangians. Note
that the full Horndeski theory (\emph{H45}) includes \emph{H3} which,
in turn, includes \emph{BD}. They all correspond to ``sub-spaces" of
different dimensionality in the space of theories, schematically shown
in the following figure.
\begin{figure}[H] \ 
\centering
   \includegraphics[scale=0.4]{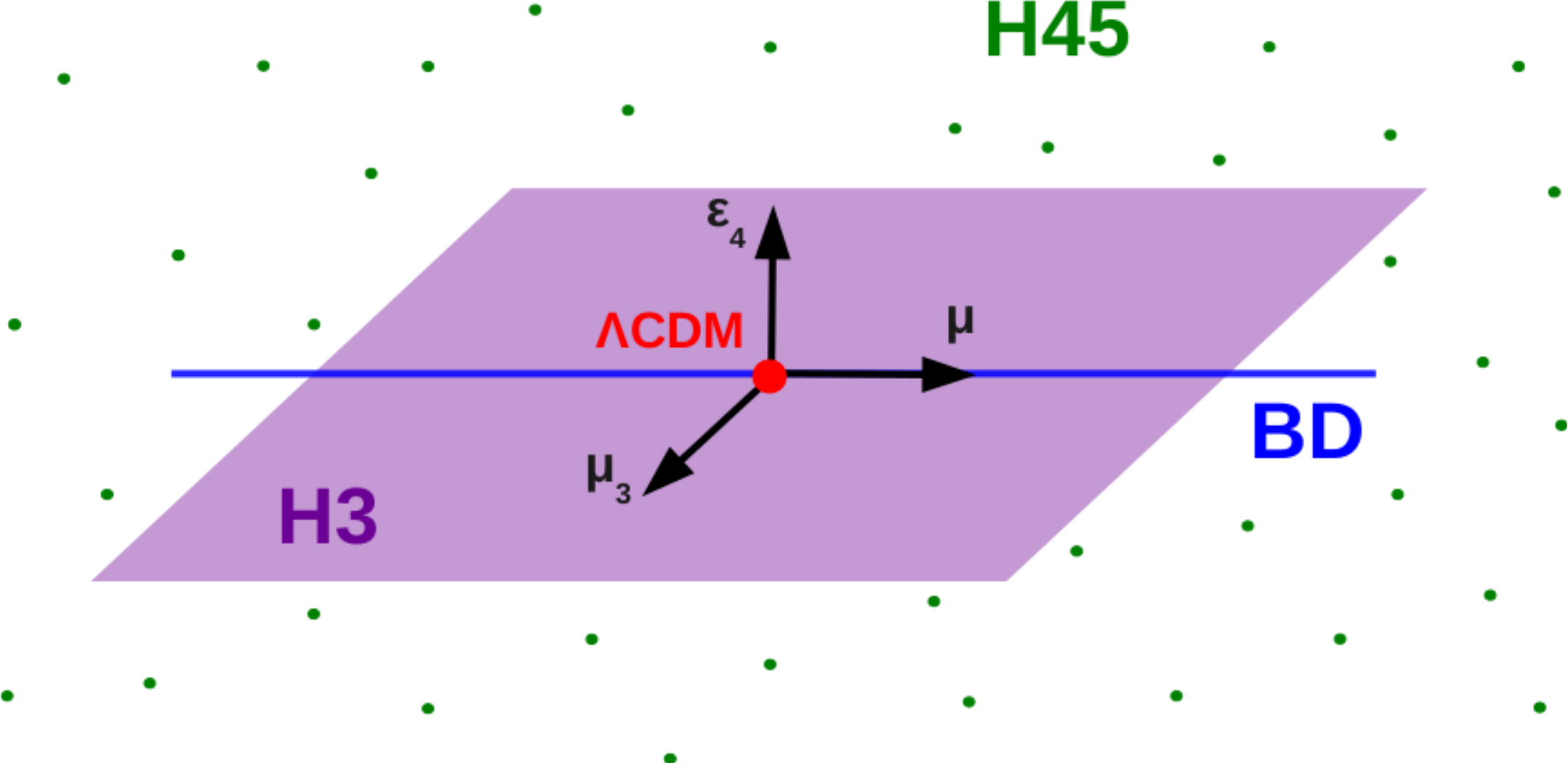}  
\caption{\label{fig:theory_space}  
Pictorial representation of the space of theories spanned by the EFT coordinates $\mu, \mu_3$ and $\epsilon_4$.}
\end{figure}

The outline of the paper is as follows.
We give an overview of the effective field theory formulation
in Sec.~\ref{Effects}, focusing on the quasi-static, sub-Hubble (or
Newtonian) limit. We then work out the expressions for the gravitational coupling(s) and
the gravitational slip parameter. We further reduce the degrees of freedom of the formalism by 
implementing  viability conditions and  by fixing the background expansion history (we choose
the  $\Lambda$CDM model which best fits the Planck's satellite  data \cite{planck}).
In Sec.~\ref{Predictions}, we introduce the parametrization of the
requisite time-dependent functions, and  systematically scan the
parameter space of general dark energy theories. We highlight 
the generic and robust predictions for the relevant
large-scale-structure observables,
such as the  linear growth rate or the gravitational slip parameter.
We conclude in Sec.~\ref{Conclusions} with a summary of the main
results, and qualitative explanations for them.

\section{Effects of dark energy on cosmological perturbations}
\label{Effects}

There is a range of scales on which extracting perturbation
observables from modified gravity (MG) theories is relatively
straightforward: the window of comoving Fourier modes $k_{\rm sh} < k
< k_{\rm nl}$. For momenta less than the non-linear scale, $k_{\rm nl} \simeq (10\, {\rm
  Mpc})^{-1}$, one can trust linear perturbation theory. For momenta well above the sound horizon scale $k_{\rm sh}
\simeq a H/c_s$ ($c_s$ is the speed of sound of dark energy
fluctuations), one can neglect the time derivatives of the metric and scalar fluctuations in the linear equations, the so called \emph{quasi-static approximation}~\cite{Noller:2013wca,Sawicki:2015zya}.  In the quasi-static regime, it is possible to compute algebraically the effective Newton constant  $G_{\rm eff}(t,k) $ and the gravitational slip parameter $\gs (t,k)$ of a given MG theory. The entire set of perturbation equations then reduces to
\begin{align}
-\frac{k^2}{a^2} \Phi &= 4 \pi G_{\rm eff}(t, k) \delta \rho_m \,, \label{poisson}\\
\gs(t,k) &= \frac{\Psi}{\Phi},  \, \label{slipgrav}
\end{align}
where we have adopted the following convention for the perturbed metric in Newtonian gauge,
 \begin{equation}\label{newtonian}
ds^2 = - (1+ 2 \Phi) dt^2 + a^2 (1 - 2 \Psi) \delta_{i j} d x^i d x^j\, .
\end{equation}
Eqs.~\eqref{poisson} and~\eqref{slipgrav} should be supplemented by
the equations for the matter fluctuation $\delta\rho_m$. 
Note that we work in the  ``Jordan frame"---one where all matter
fields are minimally coupled to the metric 
(see~\cite{Gleyzes:2015pma} for a relaxation of this
assumption within the effective field theory formalism)
and do not couple directly to the dark energy fluctuation.
This is the frame of most direct physical
interpretation~\cite{Nitti:2012ev}, 
where bodies follow geodesics\footnote{Geodesic motion can be violated in certain theories that exhibit screening such as in chameleon theories, though this assumption remains valid for the kind of screening found in the galileons~\cite{HuiNicolisStubbs}.}.

In the rest of this section, we derive $G_{\rm eff}$ and $\gs$  in the framework of the effective field theory of dark energy (EFT of DE). 
We also clarify the relation between $\geff$ and the Newton constant $G_{N}$ measured, for example, in Cavendish experiments. 
Finally, we summarize the viability conditions to be imposed on our parameter space. 

\subsection{Gravitational couplings and gravitational slip}\label{observables}

The linear cosmological perturbations of the class of theories
considered in this paper are described by the unitary gauge action
given in App.~\ref{app-a}. By moving to the more practical Newtonian
gauge, we reintroduce the perturbation $\pi$ of the scalar field
(dark energy fluctuation),
along with the Newtonian potentials $\Phi$ and $\Psi$ defined
in~\eqref{newtonian}. The action quadratic in these quantities is very
involved, but can be considerably simplified by the following
approximations. First, we apply the quasi-static approximation,  and
neglect the time derivatives in the gravity-scalar sector. 
Second, in surveys of large scale structure, 
we generally observe modes well inside the Hubble horizon. 
We thus ignore mass terms in the perturbation quantities
$\pi$, $\Phi$ and $\Psi$, because they are naturally of order Hubble
and are small compared to the gradient terms---we retain
the lowest gradients in the spirit of a gradient expansion\footnote{Note that we are not exploring chameleon
and $f(R)$  models, which phenomenologically require a mass for the
scalar field much larger than $H$~\cite{WangHuiKhoury} (see also
footnote 1).}. In this  limit,  all the dark energy models containing up to one scalar degree of 
freedom, and whose equations of motion have no more than two derivatives, are described by~\cite{GLPV3}
\begin{align}
\label{actionn}
S = \int a\,  M^2&\left[(\vn \Psi)^2 - 2 (1+ \epsilon_4) \vn \Phi \vn \Psi - 2 (\mu + \mathring{\epsilon}_4) \vn \Psi \vn \pi  \right. \\ 
& \left.  + \,  (\mu - \mu_3) \vn \Phi \vn \pi  -\left({\cal C} + \frac{\mathring{\mu}_3}{2} - \dot H \epsilon_4 + H  \mathring{\epsilon}_4 \right) (\vn \pi)^2 \right] - a^3 \, \Phi \delta \rho_m, \nonumber
\end{align}
where $\delta \rho_m$ is the perturbation of the non-relativistic energy density, a dot means derivative w.r.t. proper time and $\pi$ represents the perturbation of the scalar field.   
The non-minimal coupling $\mu$ and the ``bare Planck mass" $M$ are related by
\begin{equation}\label{mu}
\mu \equiv \frac{d \ln M^2(t)}{dt}, 
\end{equation} 
and the expression of $\cal C$ is given as
\begin{align}
\cb \ &  =   \   \frac{1}{2} ( H \mu - \dot{\mu} - \mu^2 )  - \dot H -  \frac{\rho_m}{2 M^2} \;. \label{c2}
\end{align}
In the above, $\rho_m\propto a^{-3}$ is the background energy density of non-relativistic matter. In minimally coupled scalar field models, we can think of ${\cal C}$ as related to the kinetic energy density of the field, ${\cal C} \sim \dot \phi^2/M^2$. 
To simplify the notation, we have also defined with a circle some ``generalized time derivatives": 
\begin{align}
\mathring{\mu}_3  \ &\equiv \ \dot \mu_3 + \mu \mu_3 + H \mu_3 ,\\
\mathring{\epsilon}_4  \ &\equiv \ \dot \epsilon_4 + \mu \epsilon_4 + H \epsilon_4\, .
\end{align}

Variation of~\eqref{actionn} with respect to the curvature potential $\Psi$ produces the algebraic relation 
\begin{equation} \label{Psi}
\Psi = (1+\epsilon_4) \Phi + (\mu + \mathring{\epsilon}_4) \pi
\end{equation} 
that, once substituted back  into  the action, gives 
\begin{align} \label{actionn2}
S = \int a\,  M^2&\left\{-  (1+ \epsilon_4)^2 (\vn \Phi)^2 +  \left[\mu - \mu_3 - 2 (\mu + \mathring{\epsilon}_4)(1+ \epsilon_4) \right] \vn \Phi \vn \pi  \right. \\ 
& \left.   -\left[{\cal C} + (\mu + \mathring{\epsilon}_4)^2 + \frac{\mathring{\mu}_3}{2} - \dot H \epsilon_4 + H  \mathring{\epsilon}_4 \right] (\vn \pi)^2 \right\} - a^3 \, \Phi \delta \rho_m. \nonumber
\end{align}
We can solve the coupled $\pi$-$\Phi$ system by taking the variation with respect to $\pi$, giving 
\begin{equation}
\label{piPhi}
\left[2{\cal C} + 2 (\mu + \mathring{\epsilon}_4)^2 + \mathring{\mu}_3 - 2\dot H \epsilon_4 + 2 H  \mathring{\epsilon}_4 \right] \pi \ = \  \left[\mu - \mu_3 - 2 (\mu + \mathring{\epsilon}_4)(1+ \epsilon_4) \right] \Phi\, .
\end{equation}

We thus conclude that the Newtonian gravitational potential $\Phi$  satisfies the Poisson equation 
\begin{equation} \label{pi}
-\frac{k^2}{a^2}\Phi=4\pi G_{\rm eff}(t)\rho_m \delta_m\, ,
\end{equation}
where 
\begin{equation} \label{geff}
G_{\rm eff} \ = \ \frac{1}{8 \pi  M^2(1+\epsilon_4)^2} \ \frac{2 {\cb}  +  \mathring{\mu}_3  - 2 \dot H \epsilon_4 + 2 H \mathring{\epsilon}_ 4 + 2 (\mu + \mathring{\epsilon}_4)^2 \ }{\ 2 {\cb} + \mathring{\mu}_3  - 2 \dot H \epsilon_4 + 2 H \mathring{\epsilon}_ 4 + 2 \dfrac{(\mu + \mathring{\epsilon}_ 4) (\mu - \mu_3)}{1+\epsilon_4} - \dfrac{(\mu - \mu_3)^2}{2 (1+\epsilon_4)^2}  } \ .
\end{equation}

As noted, we are neglecting any possible scale-dependence of $G_{\rm eff}$.   
The $k$-dependent corrections to~\eqref{geff} become important at large distances, how large depending on the size of the mass terms that we have neglected
in~\eqref{action}. If the scalar degree of freedom, as we are
assuming, plays a relevant role in the acceleration of the Universe,
its mass is expected to be of order $H_0$ or lighter.  
Only Fourier modes approaching the Hubble scale are affected by
such mass terms.

We need now to specify the relation between the effective Newton constant $\geff$ and  the standard  Newton constant $\gn$---or, equivalently,  the Planck mass $\mp$.  Powerful Solar System \cite{Williams:2012nc,Bertotti:2003rm} and astrophysical \cite{Jain:2012tn,Vikram:2013uba} tests impose stringent limits on   modified gravity. Realistic models must incorporate ``screening" mechanisms that ensure convergence to General Relativity on small scales and/or high-density environments. Horndeski theories mainly rely on the    
Vainshtein  mechanism, which is now quite well understood also in a cosmological time dependent setup~\cite{DeFelice:2011th,Kimura:2011dc,Koyama:2013paa,Kase:2013uja}.   
In the vicinity of a massive body, the gravitational contribution of the scalar field fluctuation $\pi$ is suppressed because non-linear scalar self-interaction terms locally change the normalisation of the $\dot \pi^2$ term in the quadratic action~\cite{Babichev:2013usa}. This is equivalent to  switching off the couplings of the canonically normalized $\pi$ to the other fields $\Psi$ and $\Phi$.
Such an astrophysical effect cannot be encoded in our quadratic action~\eqref{actionn}, whose couplings depend only on time and not on space\footnote{
There's also a ``temporal" or ``cosmological" screening which screens out modified gravity
 effects at high redshifts~\cite{Chow:2009fm}: even though our EFT formulation does not
 provide a microscopic description of how this comes about, it does effectively
 account for such a temporal effect via the time-dependent functions, $\mu$, $\mu_3$ and $\epsilon_4$.}. However,  by inspection of action~\eqref{actionn2} 
 we conclude that in a screened environment, the ``bare" gravitational coupling $M^{-2}$ gets simply dressed by a factor of $(1+\epsilon_4)^{-2}$. This is what we obtain in the Poisson equation if we switch off the mixing term $\vec\nabla\Phi \vec \nabla \pi$. We thus define a \emph{screened} gravitational coupling 
 \begin{equation}
 G_{\rm sc}(t) \ \equiv \ \frac{1}{8 \pi M^2 (1+\epsilon_4)^2}\, ,
 \label{gsc}
 \end{equation} 
 valid only in (totally) screened environments. 
 
 Since we live and perform experiments in a screened environment, the value of $G_{\rm cs}$ evaluated today is the Newton constant measured for example by Cavendish experiments, 
\begin{equation} \label{gn}
\gn \ \equiv\ \frac{1}{8 \pi  \mp^2} \ \simeq \ \frac{1}{8 \pi  M^2(t_0)[1+\epsilon_4(t_0)]^2} \, .
\end{equation}

To complete the discussion, we would like to mention that the coupling of gravitational waves to matter contains, instead, only one factor of $(1+\epsilon_4)$ at the denominator, and not two as $G_{\rm sc}$~\cite{GLPV,GLPV3}. In ref.~\cite{Sawicki}, the corresponding mass scale has been associated with the ``Planck mass". On the basis of the above arguments, here we find it more natural to define $\mp$ as in~\eqref{gn} instead.

In summary, beyond the ``bare" mass $M$ multiplying the Einstein Hilbert term in the unitary gauge action~\eqref{action}, we can define three gravitational couplings: 
\begin{framed}
\begin{itemize}
\item $G_{\rm gw} \equiv \frac{1}{8 \pi M^2 (1+\epsilon_4)}$ is the coupling of gravity waves to matter.
\item $G_{\rm eff}$ given in eq.~\eqref{geff} is the gravitational coupling of two objects in the quasi-static approximation and in the linear regime, \emph{i.e. when screening is not effective}. This is the quantity which is relevant  on large (linear) cosmological scales. 
\item $G_{\rm sc} \equiv \frac{1}{8 \pi M^2 (1+\epsilon_4)^2}$ is the gravitational coupling of two objects in the quasi-static approximation \emph{when screening is effective}. Since the solar system is a screened environment, this is also the Newton constant measured by a Cavendish experiment, once evaluated at the present day: $\gn = G_{\rm sc}(t_0)$.
\end{itemize}
\end{framed}

Finally we note that  even in the absence of anisotropic stress, the scalar and gravitational perturbations $\Phi$ and $\Psi$ are not anymore of equal amplitude, 
as predicted by  general relativity.  It is customary to describe deviations from the standard scenario by defining  the gravitational slip parameter 
$\gs \ \equiv \ \frac{\Psi}{\Phi}$, which can be easily expressed  within our formalism.
By  using~\eqref{Psi} and~\eqref{pi} we obtain 
\begin{equation} \label{postn}
\gs = 1 - \frac{(\mu + \mathring{\epsilon}_ 4) (\mu + \mu_3 + 2  \mathring{\epsilon}_ 4) - \epsilon_4 (2 \cb +  \mathring{\mu}_3 - 2 \dot H \epsilon_4 + 2 H \mathring{\epsilon}_ 4)}{ 2 \cb +  \mathring{\mu}_3 - 2 \dot H \epsilon_4 + 2 H \mathring{\epsilon}_ 4 + 2 (\mu + \mathring{\epsilon}_ 4)^2}\, .
\end{equation}
As for $\geff$, we note that not all deviations of $\gs$ from 1 are
screenable in the conventional sense. From eq.~\eqref{Psi}, we see
that in the presence of a non-vanishing $\epsilon_4$, the Newtonian potentials $\Psi$ and $\Phi$ are detuned from each other, even in environments where the $\pi$ fluctuations are heavily suppressed, $\eta_{\rm sc} \simeq 1+\epsilon_4$.

The above dark energy observables, $\geff$ and $\gs$, depend on  the six  functions $H(t)$, $M(t)$, $\mathcal{C}(t)$, $\mu (t)$, $\mu_3 (t)$ and $\epsilon_4 (t)$, which are constrained by Eqs.~\eqref{mu} and~\eqref{c2}. Note also that  the coupling $\mu_2 (t)$ does not appear explicitly  in the observables,  it only plays a role in the stability conditions and the speed of sound of dark energy, as we show in the following. Below, we discuss how to reduce the dimensionality of this functional space to 
three constant parameters and three functions. 

\subsection{Theoretical constraints: viability conditions}

One would expect a theory such as the EFT of DE, which depends on five free functions of time, to be virtually unconstrained and therefore un-predictive.
What we will show instead is that, one can indeed bound the time
evolution history of the relevant DE 
quantities such as $\geff$ and $\gs$ defined above. 
The key is to demand that the DE theories be free of physical
pathologies.   

We demand that a healthy DE theory satisfies the following four
conditions: it must not be affected by ghosts, or by gradient
instabilities; the scalar  as well as tensor perturbations must
propagate at luminal or subluminal speeds~\cite{Adams:2006sv}. In what
follows we simply refer to all these criteria as the stability or
viability conditions. 
These conditions must be satisfied at any time $t$ in the past, while we do not enforce their future validity.
We collect here the algebraic relations that enforce these
requirements, allowing us to bound the time dependent couplings. 
An in-depth derivation and discussion of these relations  can be found in paper I.

\begin{align}\label{conditions}
A  \ &> \ 0 \qquad \qquad \text{no-ghost  condition},\\[2mm] \label{gradcon}
B  \ &\ge \ 0 \qquad \qquad \text{gradient-stability  condition},\\[2mm] \label{sublum}
c_s^2\ =\ \frac{B}{A} & \leqslant \ 1  \qquad \qquad \text{scalar subluminarity  condition},\\[2mm] 
c_T^2\ =\dfrac{1}{1+\epsilon_4} &\leqslant \ 1  \qquad \qquad \text{tensor subluminarity  condition}, \label{lastcon}
\end{align}
where we have defined
\begin{align} \label{A}
A \ &= \ (\cb + 2 \mu_2^2)(1+ \epsilon_4) + \frac34 (\mu-\mu_3)^2 \, , \\ \label{B}
B \ &= \ (\cb +  \frac{\mathring{\mu}_3}{2} -  \dot H \epsilon_4 +  H \mathring{\epsilon}_ 4)(1+ \epsilon_4) - (\mu - \mu_3)\left(\frac{\mu - \mu_3}{4(1+ \epsilon_4)} - \mu -  \mathring{\epsilon}_ 4\right)\, . 
\end{align}

\subsection{Observational constraints:  background expansion history}\label{background}

Recent observations tightly constrain the homogeneous background expansion history of the universe -- equivalently, its Hubble rate $H(z)$ as a function of the redshift -- to that of a spatially flat $\Lambda$CDM model~\cite{Betoule:2014frx,planck,Aubourg:2014yra}.  
We thus assume 
\begin{equation} \label{h}
H^2(z) = H_0^2\left[x_0 (1+z)^3 + \left(1-x_0\right)(1+z)^{3(1+w_{\rm eff})}\right]\, .
\end{equation}
The quantities $x_0$---the present fractional matter density of the
background -- and $w_{\rm eff}$---the effective equation of state
parameter -- are free parameters, though observations suggest
$x_0$ and $w_{\rm eff}$ must be close to $0.3$ and $-1$ respectively.
Since we are interested in the recent expansion history, we have neglected the contribution of radiation. 
  
The fractional matter density of the background reference model calculated at any epoch, $x$, proves a useful time variable for late-time cosmology, smoothly interpolating between $x=1$, deep in the matter dominated era, and its present value $x_0 \simeq 0.3$. Its expression as a function of the redshift is
\begin{equation}
\label{xdef}
x\ = \ \frac{x_0}{x_0 + (1- x_0) (1+ z)^{3 w_{\rm eff}}}\, .
\end{equation}

Having specified the background geometry $H(z)$, let us now see how this influences the functions $M^2(t)$ and $\cb(t)$ that are needed to compute $\geff$~\eqref{geff} and $\gs$~\eqref{postn}.  
First of all, note that $M^2(t)$ is not an independent free function
of the formalism but is related to the non-minimal coupling $\mu$. 
By inverting eq.~\eqref{mu} we obtain
\begin{equation}\label{M}
M^2(t) \ = \ \frac{\mp^2}{(1+\epsilon_4^0)^2} \, \exp\left(\int_{t_0}^t dt' \mu(t')\right),
\end{equation}  
where the initial conditions have been set according to~\eqref{gn}. 
In a similar way, the evolution equations for the background (see \emph{e.g.}~\cite{EFTOr,pheno}) result in the expression of $\cal C$ given in~\eqref{c2}. There, 
$\rho_m$ represents the \emph{physical} energy density of non-relativistic matter. By ``physical" we mean the quantity appearing in the energy momentum tensor. It scales as $a^{-3}$ since we are in the Jordan frame and $p_m\simeq 0$. With this quantity, we can define the physical fractional energy density 
\begin{equation}
\Omega_m^0 \equiv \frac{\rho_m(t_0)}{3 \mp^2 H_0^2}\, .
\end{equation}
In principle, one could try to measure $\rho_m(t_0)$ by directly
weighing the total amount of baryons and dark matter, for instance
within a Hubble volume. It is worth emphasizing that, in theories of
modified gravity, $\Omega_m^0$ needs not be the same\footnote{One might
  argue that, since the background evolution is anyway degenerate
  between the two dark components, the distinction between $x_0$ and
  $\Omega_m^0$ is merely academic. It is true that, in the Friedmann
  equations, we can always relabel some fraction of dark matter as
  ``dark energy", but at the price of changing the equation of state
  of the latter. By assuming that $w_{\rm eff}$ be constant---and, in
  the rest of the paper, $w_{\rm eff} = -1$ in particular---we break
  the dark-sector degeneracy. Therefore, the distinction between $x_0$
  and $\Omega_m^0$ is in principle important.} as $x_0$.
 The latter is a purely geometrical quantity, a proxy for the behavior of $H(z)$. It proves useful to define, as in paper I \footnote{Note the change of notation with respect to Ref.~\cite{pheno}. $\Omega_m^0$ defined here \emph{is not} the present value of the function $\Omega_m$ defined there. On the other hand, $\kappa$ represents the same quantity.},
\begin{equation}\label{kappa}
\kappa\ \equiv \ \frac{\Omega_m^0}{x_0}\, .
\end{equation}

In conclusion, the behavior of $\geff$ and $\gs$ is completely specified once the  three parameters $x_{0}, \ w_{\rm eff}$, and  $\kappa$,
and the three functions,  $ \mu(t), \ \mu_3(t),\ \epsilon_4(t)$, are supplied. Effectively, these functions are coordinates in the parameter space of modified gravity theories.

\section{General predictions} 
\label{Predictions}

In this section we describe the redshift scaling of  interesting cosmological observables such as  the effective Newton constant $\geff$, the linear growth rate of large scale structure $f$ and the gravitational slip parameter $\gs$. 

\subsection{Methodology}

Before presenting our results,  let us  summarize  the  framework within which our predictions are derived. 
First of all,  we enforce that viable modified gravity  models reproduce the $\Lambda$CDM background history of the universe. In other words,  we choose  as  background $H(z)$ that of a flat $\Lambda$CDM universe with parameters set by Planck ($x_0=0.314$ and  $w_{\rm eff}=-1$).

As for the  perturbed sector of the theory,  we make the following assumptions

\begin{itemize}

\item We adopt  $x$ as a useful time variable interpolating between the matter dominated era ($x =1$) and today ($x = x_0 \simeq 0.3$), and  expand the  coupling functions  up to second order  as follows:
\begin{align} \label{tay1}
\mu\left(x\right)\ &=\ H(1-x) \left[\pz_1+\pu_1 \left(x -x_0\right)+\pd_1 \left(x -x_0\right)^2 \right ]\, ,\\[2mm]
\mu_3\left(x\right)\ &=\ H(1-x) \left[\pz_3+\pu_3 \left(x -x_0\right)+\pd_3 \left(x -x_0\right)^2 \right ]\, ,\label{tay2}\\[2mm]
\epsilon_4\left(x\right)\ &=\ \ \ (1-x) \, \left[\pz_4+\pu_4\left(x -x_0\right)+\pd_4 \left(x -x_0\right)^2 \right ]\, \label{tay3}.
\end{align}
On the other hand, we set $\mu_2 = 0$  since this function's only
effect is that of lowering the speed of sound of dark energy (see
eq.~\ref{sublum}) and thus reducing the range of validity of the
quasi static limit, an approximation used in the present analysis. 
Secondly, we want to recover standard cosmology at very early
times. This is why the Taylor expansions are multiplied by an overall
factor of $(1- x)$ which serves to switch off the couplings at early
epochs. This ties the modification of gravity to the (recent)
phenomenon of dark energy. 
\item In the same spirit, we demand that the function ${\cal C}$, appearing in $\geff$ and $\gs$, grows at early epochs less rapidly than $H^2$ {\it i.e.}  ${\cal C}/H^2 \rightarrow 0$ at high redshift. In the limit of an (uncoupled) quintessence field $\phi$, ${\cal C}$ is nothing else than the kinetic fraction of the energy density of the field, ${\cal C} \sim \dot \phi^2/\mp^2$. We are thus requiring that dark energy be subdominant at high redshift, which is usually the case in explicit models~\cite{Chow:2009fm}. By inspection of equation~\eqref{c2}, we see that we need a cancellation  between the last two terms on the RHS. By substituting $\rho_m = 3 \mp^2 \kappa x H^2$ and using~\eqref{kappa}, we see that this condition is equivalent to setting the value of $M^2$ at early times,\footnote{This is the same relation found in paper I with a slightly different reasoning.}
\begin{equation} \label{limitM}
M^2\ \underset{x \rightarrow 1}{\longrightarrow} \ \kappa \mp^2\, .
\end{equation}
By using~\eqref{M}, this imposes a constraint on the Taylor coefficients of $\mu$:
\begin{equation} \label{constrain}
 \ln\left[\kappa(1+\epsilon_4^0)^2\right] = 
\frac{1-x_0}{6 w_{\rm eff}} \left[2 \pu_1+\pd_1\left(1 -3 x_0 \right)\right]-\frac{\ln \left(x_0\right)}{3 w_{\rm eff}}\left[\pz_1 - x_0 \pu_1 + x_0^2 \pd_1\right] .
\end{equation}
As we point out in the rest of the paper, some of the general features that we find are indeed related with this requirement.

\item All of our analysis is carried out by assuming $\kappa=1$, \emph{i.e.}, by requiring that  the physical and the effective reduced matter densities coincide at the present time.  This choice is somewhat suggested by observations. Direct observations of the mass-to-light ratio on large scales lead to a value of $\Omega_m^0$ (see e.g.~\cite{Bahcall:2013epa}) that is compatible with the ``geometrical quantity" (here called $x_0$) obtained by fitting cosmological distances.

\item We explore the space of non-minimal couplings $\mu$, $\mu_3$ and
  $\epsilon_4$ that covers the entire set of Hordenski theories by
  randomly generating points in the  8-dimensional parameter space
  of coefficients $p_n^{(i)}$ (note that in the $\mu$-sector the three
  $p_1^{(i)}$ are related by eq.~\ref{constrain}). We reject those
  points that do not pass the stability
  conditions~\eqref{conditions}-\eqref{lastcon} until we have produced
  $10^4$ viable models. Note that, when dealing with the full
  8-dimensional volume of parameters, the chance of hitting a stable
  theory is much lower than $1 \%$. Because we want the quasi-static
  approximation to have a large enough range of applicability, we also
  impose $c_s>0.1$, which allows us to cover the Fourier volume of the
  EUCLID mission~\cite{Sawicki:2015zya}. We emphasize,
  however, that this is a very weak selection effect on our randomly generated models.

\item There is no ``natural unit" in the space of Taylor coefficients. Mostly, we randomly generate them in the interval $p_n^{(i)} \in [-2,2]$,
so that the relevant observables such as  $f \sigma_8$ (see Figure 4) 
span the interval of uncertainty of current data. We have checked the
robustness of our findings by expanding the ranges for different
coefficients, and by raising  the order of the expansions. A
sample of these checks is reproduced in
Figure~\ref{fig:fhisto}. Additional checks that we have done include
modeling the EFT couplings as step functions of variable height and
width at characteristic redshifts and using different parameterizations for the \emph{BD} sector. In particular, the one used in paper I is illustrated in App. \ref{wparam}. 

\end{itemize}

\subsection{Evolution of the effective Newton constant}
\begin{figure}[t!] \vspace{-.6cm}
\centering
   \includegraphics[scale=0.85]{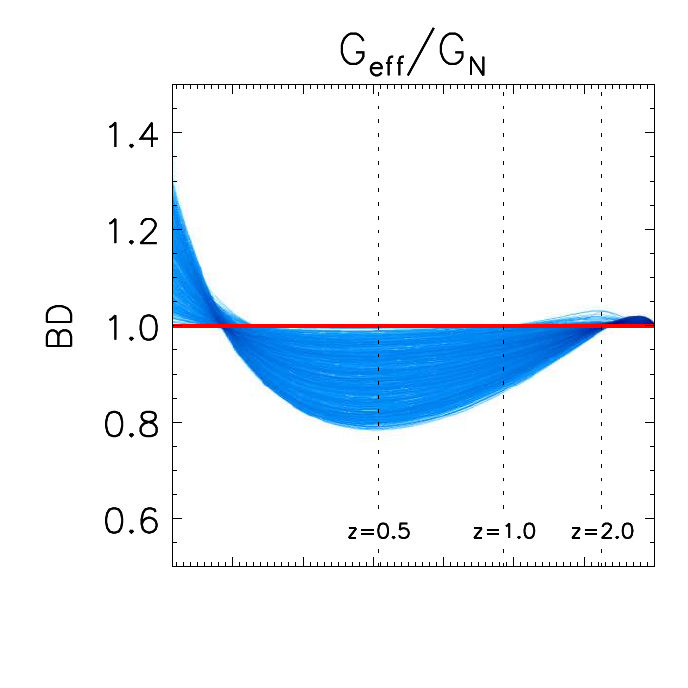}  \hskip-17mm
   \includegraphics[scale=0.85]{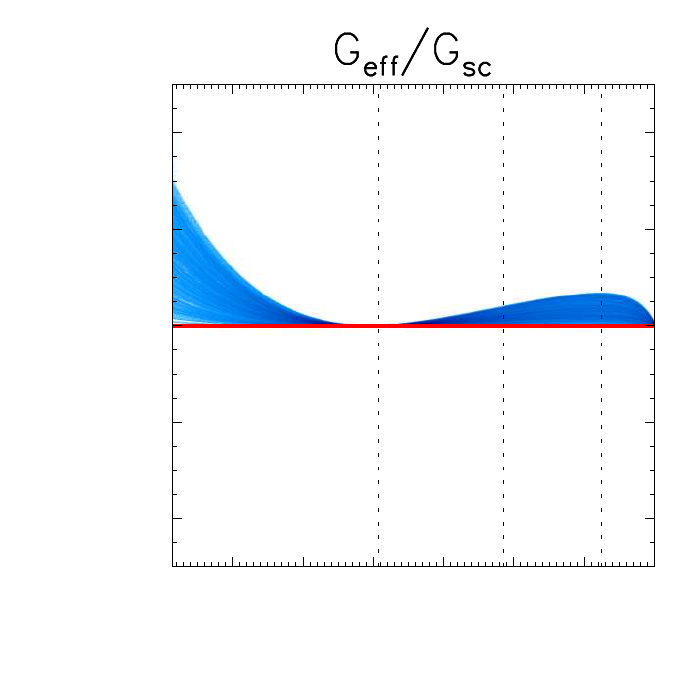} \hskip-17mm
   \includegraphics[scale=0.85]{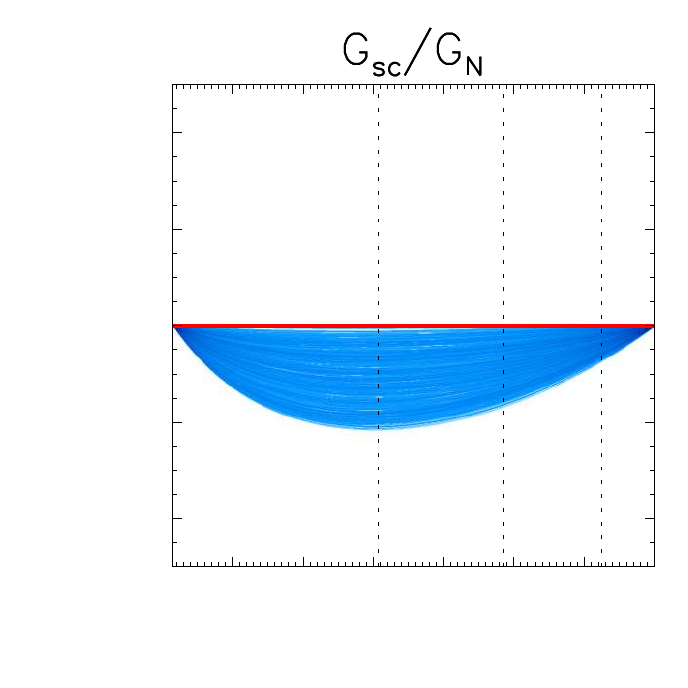}   \vskip-16mm
   \includegraphics[scale=0.85]{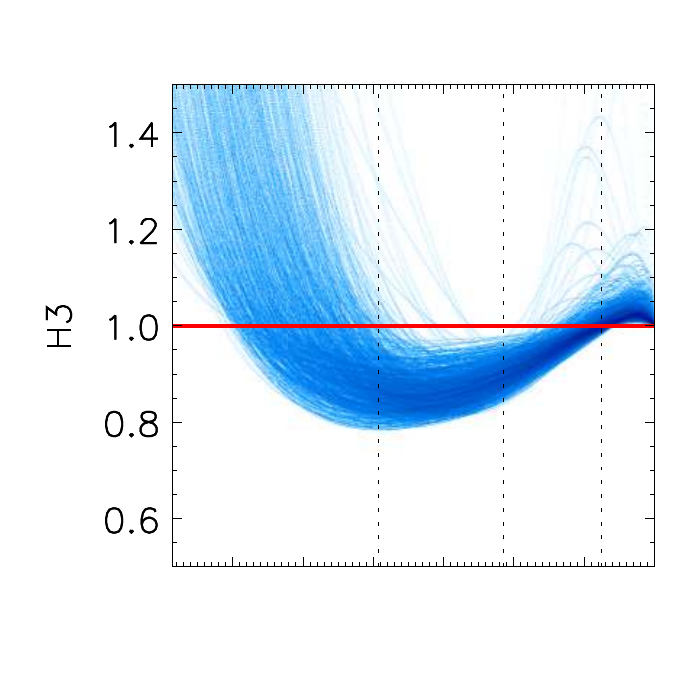}  \hskip-17mm
   \includegraphics[scale=0.85]{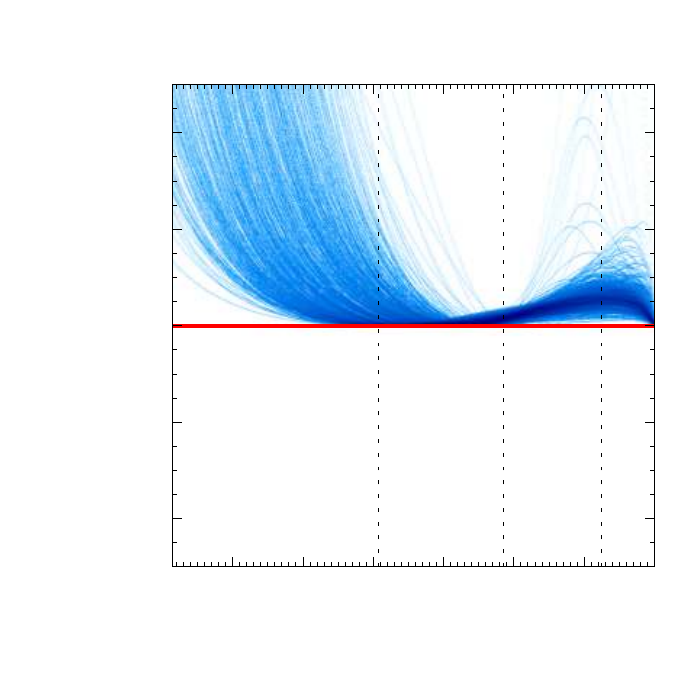} \hskip-17mm
   \includegraphics[scale=0.85]{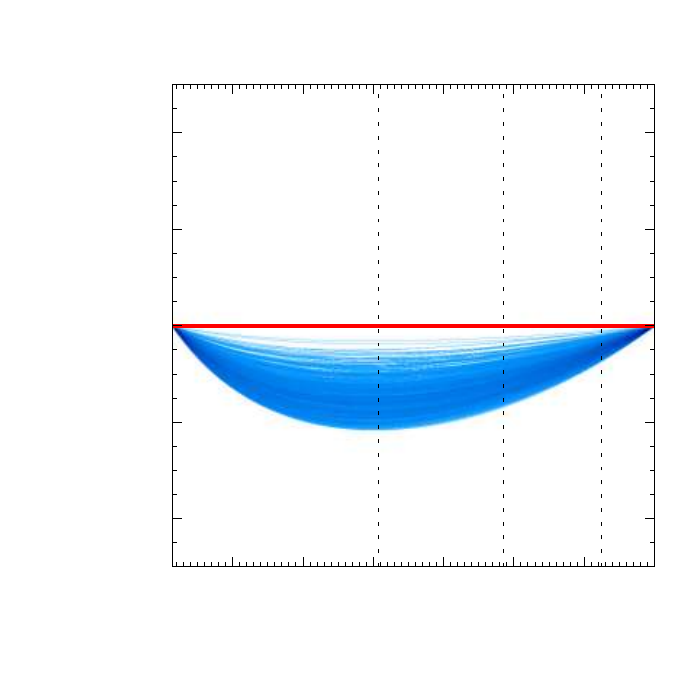}   \vskip-16mm
   \includegraphics[scale=0.85]{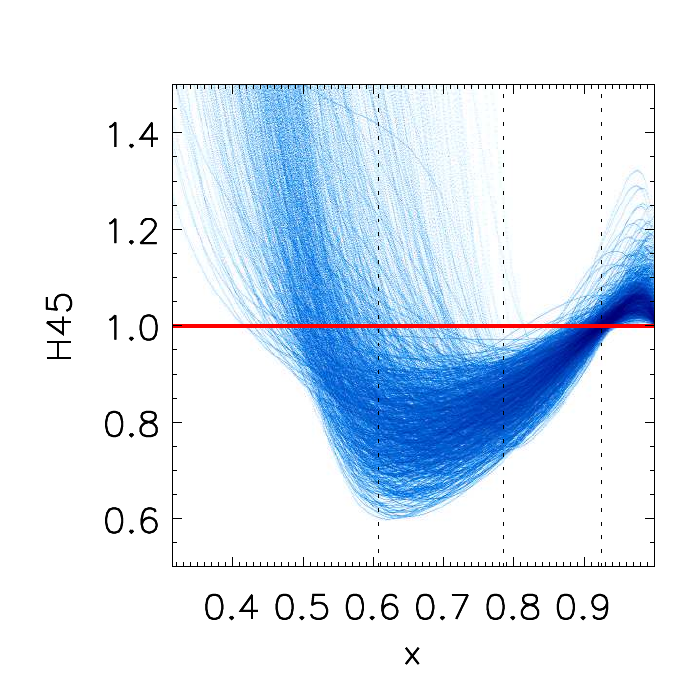}   \hskip-17mm
   \includegraphics[scale=0.85]{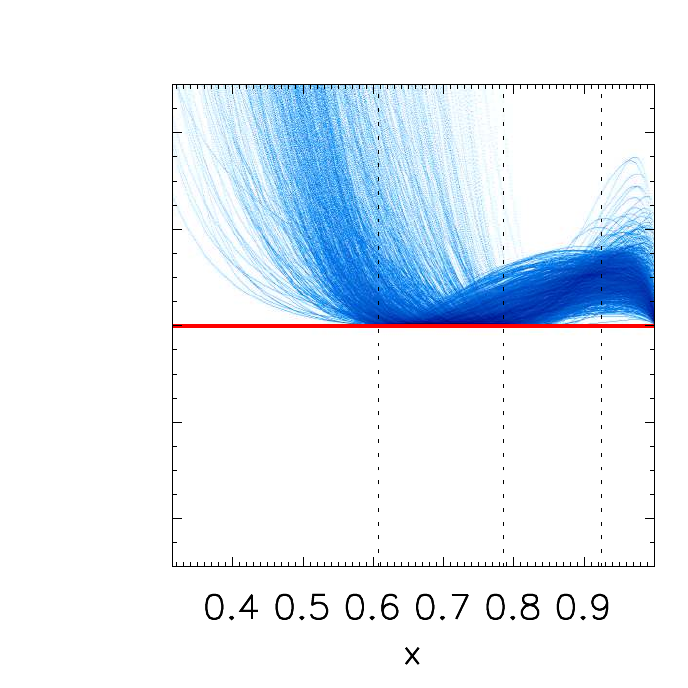}  \hskip-17mm
   \includegraphics[scale=0.85]{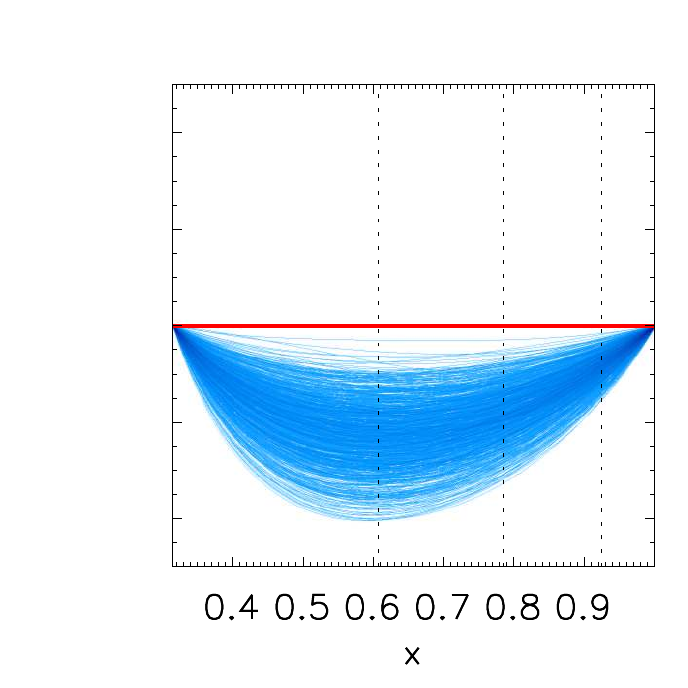}
\caption{\label{fig:geff_g3} 
 The behavior of $\geff$ (left panels) and its two separate
 contributions, from eq.~\eqref{contributions} (central and right panels),  as a function of the reduced matter density $x$ is shown for a sample of $10^3$ randomly generated models of viable \emph{BD}, \emph{H3} and \emph{H45} theories. 
The  dotted vertical lines identify, from left to right, the cosmic
epochs $z=0.5$, $z=1$ and $z=2$. The thick red line represents
$\Lambda$CDM. Note that \emph{H3} and \emph{H45} do not seem to
approach this limit in a continuous way. The point is that  \emph{BD},
\emph{H3} and \emph{H45} correspond to ``subspaces" of progressively
higher dimensions in the theory space (see figure in
Sec.~\ref{sec_1}). $\Lambda$CDM represents a well-defined limit for
all models, but the volume of stable theories for \emph{H3} and
\emph{H45} asymptotically reduces to an ``hyperplane" of lower
dimensions in the vicinity of $\Lambda$CDM. As a result,  the
neighborhood of $\Lambda$CDM is effectively a volume-zero subset for
our random generator of theories. The Monte Carlo procedure does
recover $\Lambda$CDM as a limiting case of \emph{BD} nonetheless.}
\end{figure}

As mentioned, the redshift dependence  of the effective Newton constant appears to be rather constrained by the 
stability conditions. Indeed,  Figure \ref{fig:geff_g3} shows that  the effective Newton constants  in our  large class of theories  have  similar evolution 
pattern over time.  The universal behavior of $\geff(x)$ is best
captured by the following  few features. First, all the curves display
a negative derivative at $x=1$, which implies stronger
gravitational attraction ($\geff\ge\gn$)  at early epochs ($z>2$).  This behavior was proved analytically in paper I for a restricted set of models in which the Taylor expansion \eqref{tay1}-\eqref{tay3} is retained only at zeroth order. By inspection, we see here that the effect is still there at any order in the expansion.
The amplitude of such an initial bump in $\geff(x)$ varies,  with a few models displaying also  conspicuous departures from the standard model. Also the width of the time interval over which  this early stronger gravity period extends is quite model dependent. We find that the bump can be somewhat leveled out by giving up the requirement that DE be subdominant at high redshift (eqs.~\ref{limitM} and~\ref{constrain}).  

Even more interesting is the fact that all the models consistently predict that the amplitude of the effective gravitational coupling is suppressed $\geff\leq\gn$
in the redshift range $0.5\lesssim z \lesssim 1$ before turning stronger,  once again,  at around the present epoch. 
The reason for the intermediate range suppression is that at those redshifts the dominant contribution to the total of $G_{\rm eff}$ is given by the screened gravitational coupling $G_{\rm sc}$. Stability conditions always make $G_{\rm sc}$ lower than $G_{N}$ during the whole evolution. 
The characteristic {\it S-shape}  pattern shown in Figure \ref{fig:geff_g3} is common to all models and does not depend on the degree of the Taylor expansion adopted for the coupling functions~\eqref{tay1}-\eqref{tay3}.   

 We should mention that, within covariant Galileon theories---a
 subclass of the models considered here---the same qualitative
 behavior of $\geff$ was found (see \emph{e.g.}
 Ref.~\cite{Neveu:2013mfa}, Fig. 9, Ref.~\cite{Barreira:2014}, Fig. 3), although the background evolution
 in that more constrained case is different from
 $\Lambda$CDM. Regarding the weaker gravitational attraction at
 intermediate redshifts, our results are in agreement with those of
 the recent paper~\cite{Tsujikawa:2015mga}. Less clear to us is the
 role played by the gravitational wave speed $c_T$ (related to
 $\epsilon_4$, in our language) -- the author
 of~\cite{Tsujikawa:2015mga} claims $c_T < 1$ is important for
 obtaining a weaker gravitational attraction. We find that a weaker
 gravitational attraction can be---and, in most cases, is---achieved also 
in theories (\emph{BD}, \emph{H3}) where $c_T=1$
(see also \cite{Pettorino}).

A way to make sense of why the effective gravitational constant is
stronger/weaker than the corresponding standard model value at
characteristic cosmic epochs, 
is to decompose its amplitude into two distinctive parts. We can think of $\geff/G_{N}$  as
the product of two terms, 
\begin{equation} \label{contributions}
\frac{\geff}{G_{N}}\ = \ \frac{\geff}{G_{\rm sc}}\ \frac{G_{\rm sc}}{G_{N}},
\end{equation}
which can be expressed as (see eqs.~\ref{geff} and~\ref{A}) 
\begin{equation} \label{factors}
\dfrac{\geff}{G_{\rm sc}} =1+\dfrac{1+\epsilon_4}{B} \left(\dfrac{\mu -\mu_3}{1+\epsilon_4}-(\mu +\mathring{\epsilon}_ 4) \right)^2\, , \quad \qquad
\dfrac{G_{\rm sc}}{\gn}  =  \dfrac{\mp^2}{ M^2(t)(1+\epsilon_4)^2}\; , 
\end{equation}
respectively. Stability conditions~\eqref{gradcon} and~\eqref{lastcon} imply $\geff/G_{\rm sc} \geqslant 1$. Physically, this means that the scalar field contribution to the gravitational interaction is always attractive, as expected from a (healthy) spin-0 field.   This circumstance is displayed in the second column of Figure~\ref{fig:geff_g3}.

The behavior of $G_{\rm sc}/G_{N}$, the last column of
Fig.~\ref{fig:geff_g3}, also has a physical interpretation related to the stability of the models, although somewhat more subtle.  First, note that the value of such a quantity today is unity by definition---as we have argued in~\eqref{gn}, $G_N \equiv G_{\rm sc}(t_0)$--- while at early epochs ($x=1$) it is given by condition~\eqref{limitM}. On the other hand, the overall behavior of $G_{\rm sc}/G_{N}$ as a function of  time can be understood as a product of the two independent factors, $M^{-2}$ and $(1+\epsilon_4)^{-2}$. The latter  quantity is always lower than unity because
tensor perturbations are assumed to propagate at subliminal speed (see eq.~\ref{lastcon}).
 Also $M^{-2}$ decreases as a function of the redshift (i.e. backward in time) at around the present epoch. The physical reason is better understood in the ``Einstein frame"---the frame in which the metric is decoupled from the scalar---which for the background evolution simply reads $g_{\mu \nu}^{(E)}\sim M^2 (t) g_{\mu \nu}$ (see e.g.~\cite{EFTOr}). A growth of $M$ as a function of the redshift means less acceleration in the Einstein-frame, thus implying  that the observed acceleration in the physical (Jordan) frame is due to a genuine modified gravity effect (self-acceleration). Therefore, the third column of Figure~\ref{fig:geff_g3} provides a rough estimate of the amount of self-acceleration for the various randomly generated models. Curves that deviate the most from $\Lambda$CDM (the red straight line) represent models with strong  self-acceleration,  while the opposite cases 
represent models in which acceleration is essentially due to a
negative pressure component in the energy momentum tensor. Our EFT
approach covers both types of behavior in a continuous way, although
it will be interesting to understand the specific features of those
models that are truly self-accelerating~\cite{HMPP}.

\subsection{The growth rate of matter fluctuations}

The universal evolution of the effective Newton constant is expected
to result in 
a characteristic growth history for the linear density  fluctuations 
of matter.  The prediction for this  observable is obtained by solving the equation
\begin{equation}
3 w_{\rm eff}(1-x)x f'(x)+f(x)^2+ \left [   2- \frac{3}{2} \left( w_{\rm eff} (1-x) +1 \right )  \right ] f(x) =\frac{3x}{2} \kappa \frac{\geff}{\gn}\, ,
\label{lgrowth}
\end{equation}  
which approximates the true evolution in the Newtonian regime (below
the Hubble scale) and well after the initial, radiation dominated
phase of cosmic expansion. Here, $f \equiv d{\,\rm ln\,}\delta_m/d{\,\rm
  ln\,}a$, where $\delta_m$ is the matter fluctuation and $a$ is the
scale factor.

\begin{figure}[t] 
\centering
   \includegraphics[scale=0.85]{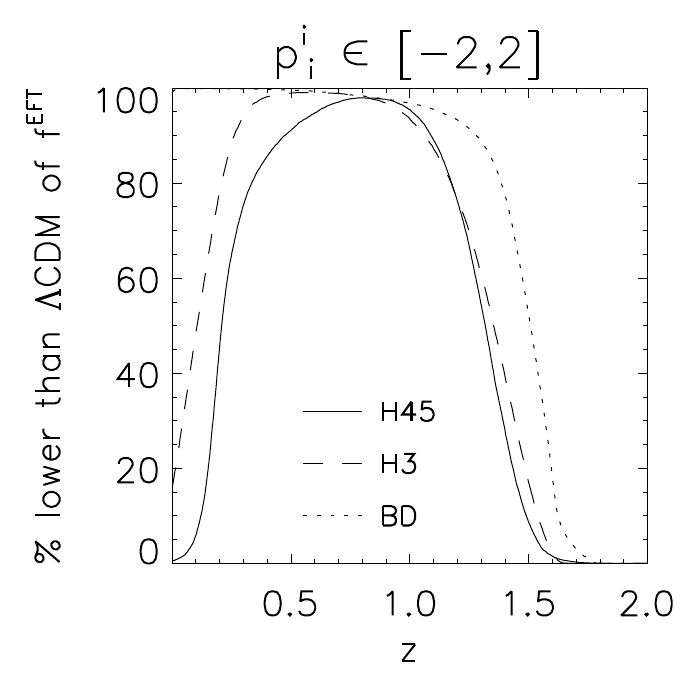}  \hskip-14mm
   \includegraphics[scale=0.85]{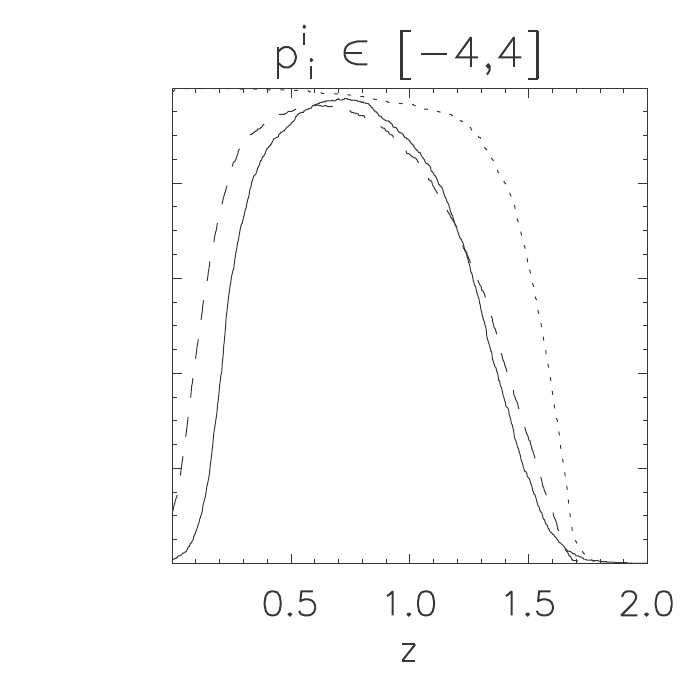}  \hskip-14mm
   \includegraphics[scale=0.85]{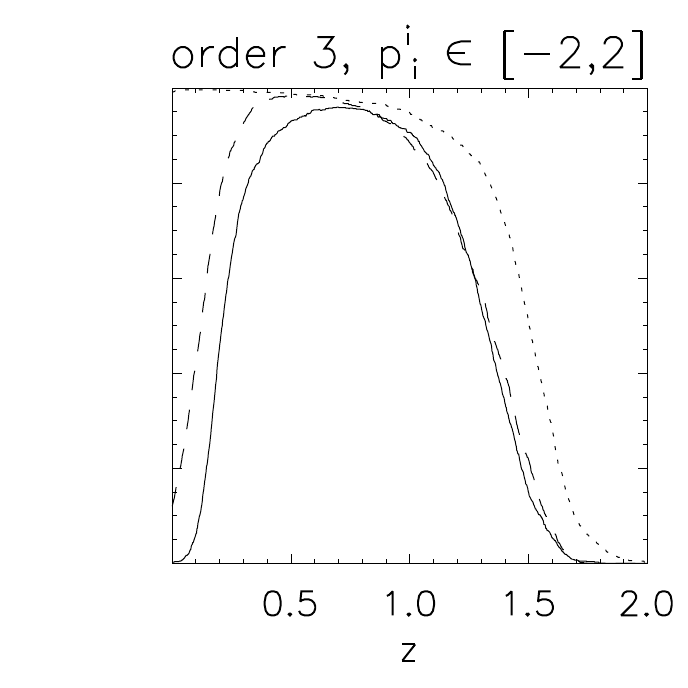} 
\caption{\label{fig:fhisto}  
Percentage of randomly generated EFT models with growth rate $f$
\emph{lower} than that predicted by $\Lambda$CDM Planck-cosmology
as a function of the redshift $z$.   We have checked the robustness of our analysis by changing the interval of the randomly generated Taylor coefficients (center) and by augmenting the order of the Taylor expansion~\eqref{tay1}-\eqref{tay3} (right).} 
\end{figure}

Figure \ref{fig:fhisto} shows that the expectation of a pattern of stronger/weaker growth phases with respect to the prediction of the standard model of cosmology
is confirmed.  Understandably, since $f$ obeys an evolution equation sourced
effectively by $x\geff$, it responds to periods of stronger/weaker
gravity with a time-lag.
Moreover, as an integrated effect, $f$ has a smaller spread compared
to $\geff$. 

The first thing worth emphasizing is that essentially all  modified gravity models with the same expansion history of $\Lambda$CDM consistently 
predict that cosmic structures grow at a stronger pace, compared to $\Lambda$CDM, at all redshifts greater than $z \sim 2$.

A second distinctive feature is that non-standard models of gravity are generally
less effective in amplifying matter fluctuations during the intermediate epochs
in which cosmic acceleration is observed, i.e. in the redshift  range  $ 0.5 \lesssim z  \lesssim 1$. Figure \ref{fig:fhisto} shows that $95\%$ of the growth rates 
predicted in the \emph{BD}, \emph{H3} and \emph{H45} classes of theories are weaker than expected in the standard $\Lambda$CDM scenario.

\begin{figure}[t] \vspace{+1cm}
\centering
   \includegraphics[scale=0.88]{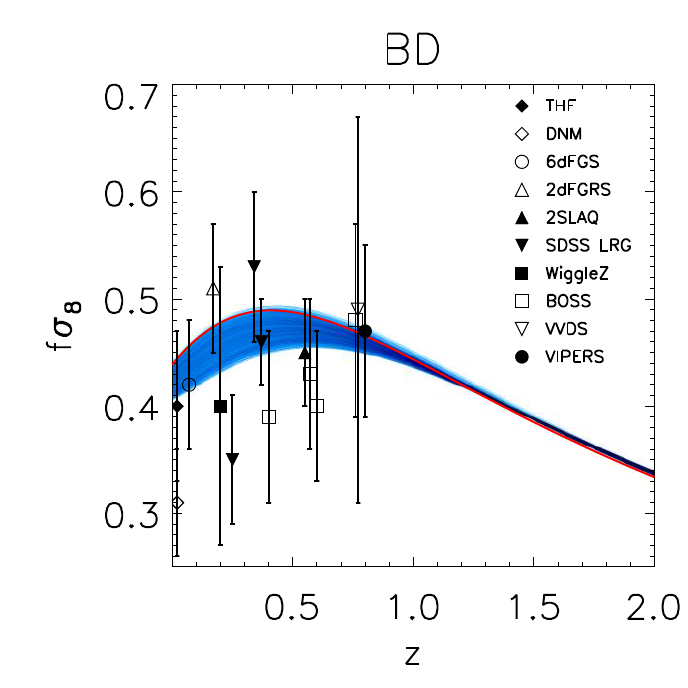}   \hskip-15mm
   \includegraphics[scale=0.88]{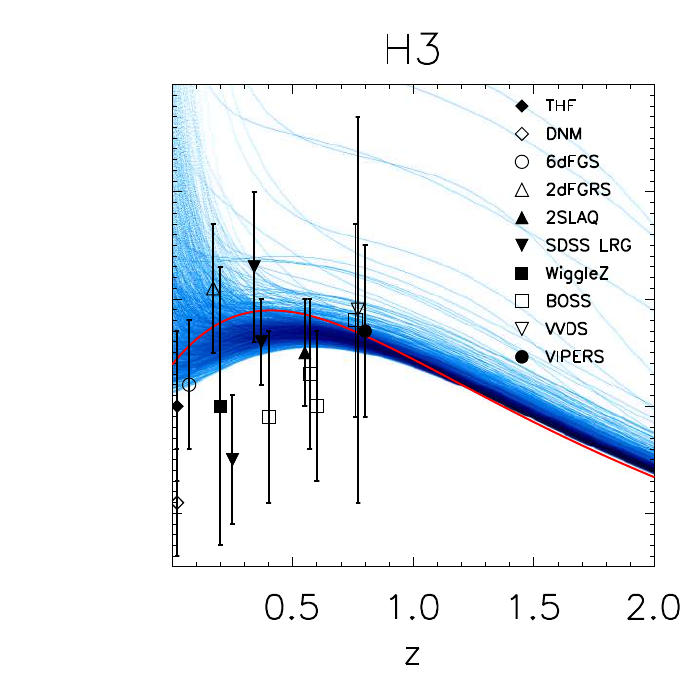}    \hskip-15mm 
   \includegraphics[scale=0.88]{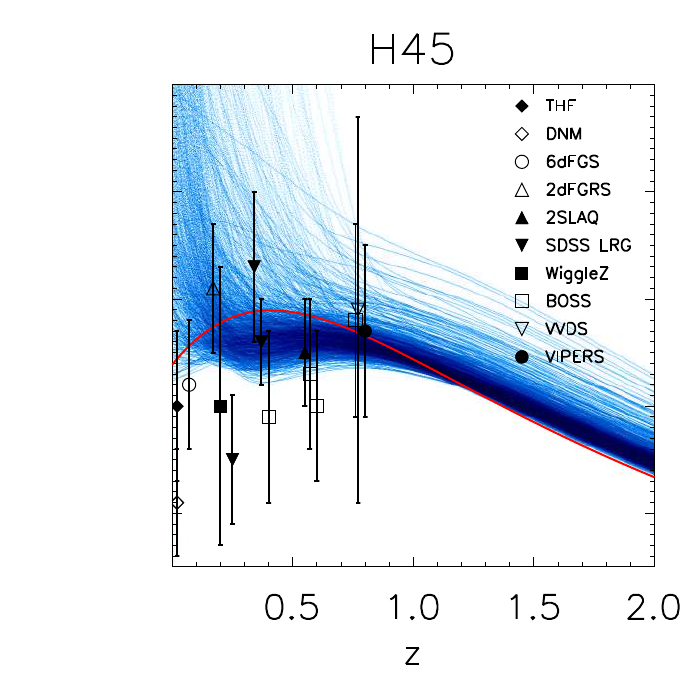}   \hskip-15mm
\caption{\label{fig:fdata}  
The redshift evolution of  $\fs$ expected in $10^3$ stable EFT models is shown and compared to data. 
Error bars represent 1$\sigma$ s.d.. The evolution predicted by the best Planck fit is shown in red.
} 
\end{figure}

These predictions can be compared with observations.
A collection of available measurements of the growth related quantity  $f \sigma_8$ is presented in Figure \ref{fig:fdata}  and compared to the most 
generic predictions of \emph{BD}, \emph{H3} and \emph{H45} theories\footnote{
We rescale the Planck value of the  present day  $\it rms$  amplitude of  matter fluctuations on a scale $8h^{-1}$ Mpc as 
$ \sigma^{EFT}_8(z)=\frac{D_{+}(z)^{EFT}}{D_{+}(z)^{\Lambda CDM}} \sigma^{\Lambda CDM}_{8}(z=0)$
where $D_{+}$ is the growing mode of linear matter density perturbations in  various dark energy models, and where the present day value 
$\sigma^{\Lambda CDM}_{8}(0)$
is set to the Planck value 0.834.}.
The current errorbars are still large, thus one should not read too
much into these plots. Nonetheless, it is intriguing that the data
suggest less growth than is predicted by $\Lambda$CDM.
If this holds up in future surveys, it would be important to check
that growth is {\it stronger} than $\Lambda$CDM at $z > 1$, as is
predicted by the bulk of the models.

\subsection{The gravitational slip parameter}

\begin{figure}[t] 
\centering
   \includegraphics[scale=0.88]{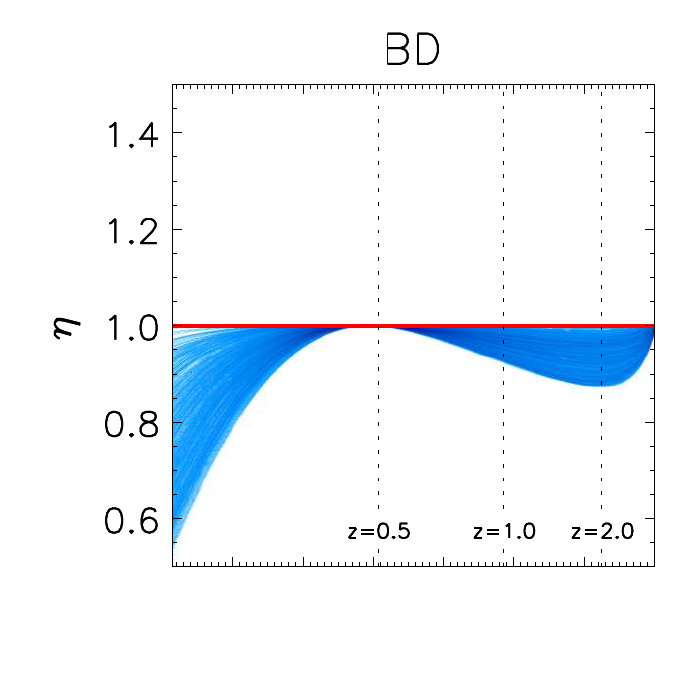}   \hskip-17mm
   \includegraphics[scale=0.88]{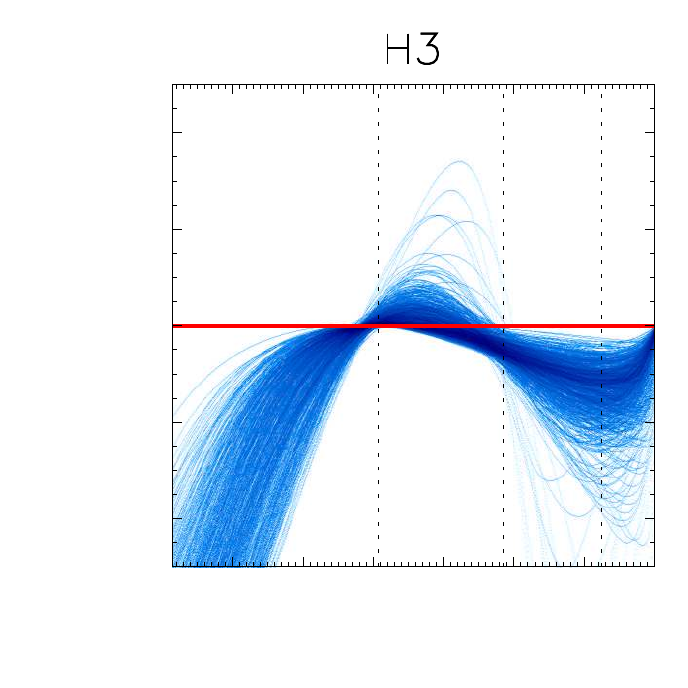}   \hskip-17mm
   \includegraphics[scale=0.88]{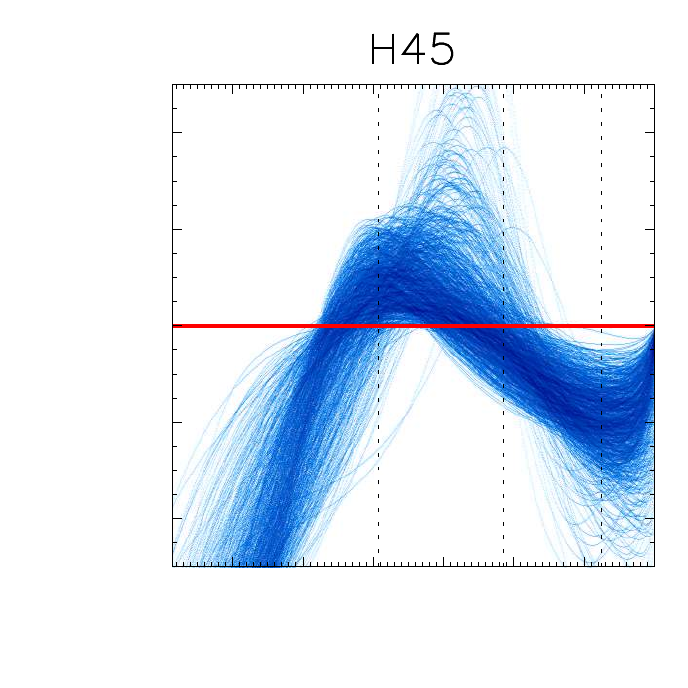}    \vskip-16mm 
   \includegraphics[scale=0.88]{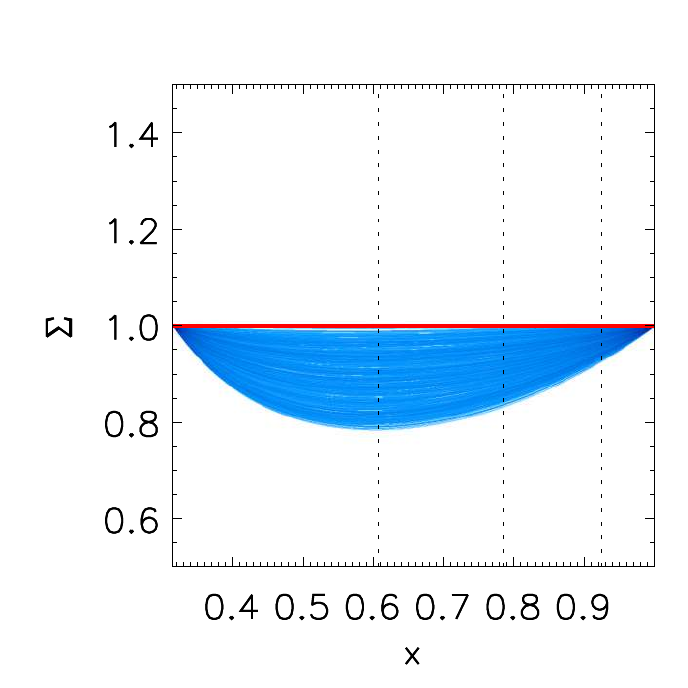} \hskip-17mm
   \includegraphics[scale=0.88]{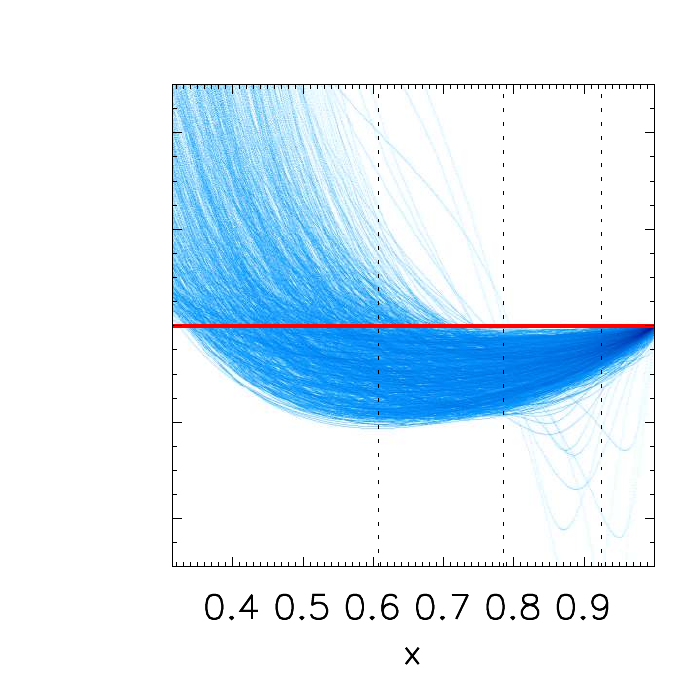} \hskip-17mm
   \includegraphics[scale=0.88]{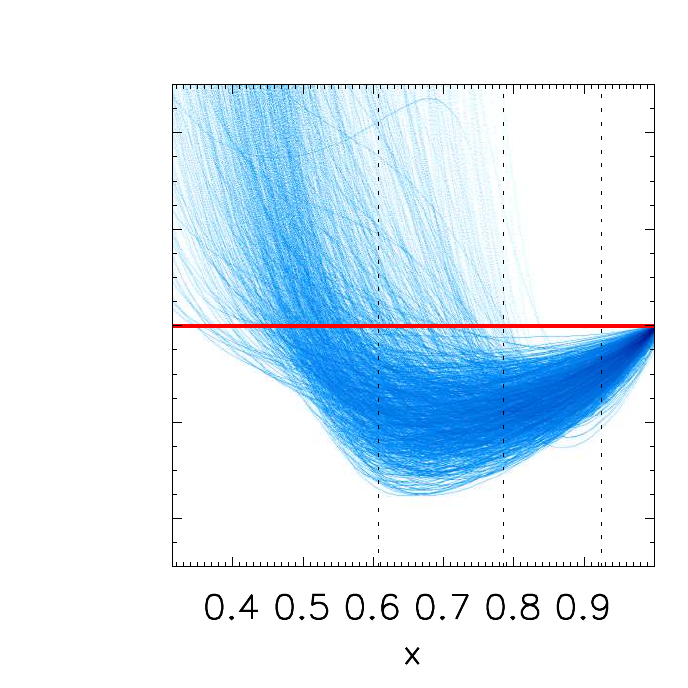}     
\caption{\label{fig:slip}  
The amplitude of the gravitational slip $\gs$ and lensing potential $\Sigma$ as a function of the reduced matter density $x$ 
for $10^{3}$ EFT models. The expected values of both these functions in a $\Lambda$CDM cosmology are shown in red.
The vertical dotted lines correspond from left to right to $z=0.5$, $z=1$ and $z=2$.} 
\end{figure}

While the peculiar velocity of galaxies falling into the large scale overdensities of matter constrains  the possible growth histories of  cosmic  structures,  
CMB and weak gravitational lensing  provide complementary probe  of
gravity, notably  they allow one to test whether   the metric potentials $\Phi$ and $\Psi$  are indeed equal, as predicted by GR   in the absence of anisotropic stress, or  differ as predicted by most non-standard models of gravity.  

Ref.~\cite{Ade:2015rim} carried out an interesting joint analysis of
current data using these three probes, obtaining constraints on two
characteristic observables that are sensitive to non-standard
behavior of the metric potentials. 
These are  the gravitational slip parameter $\gs$, already introduced  in Sec.~\ref{observables},
and the light deflection parameter or lensing potential $\Sigma$, defined in Fourier space by
\begin{equation}
-\frac{k^2}{a^2} \left ( \Psi + \Phi \right) = 8\pi \Sigma(t, k)\rho_m\delta_m\, , 
\end{equation}
which can be expressed as 
\begin{equation}
\Sigma\ =\ \frac{\geff}{\gn}\, \frac{1+\gs}{2}\, .
\end{equation}

For  \emph{BD}-like theories it is straightforward to show
analytically that the amplitude of the curvature potential $\Psi$ is
never greater than the Newtonian potential $\Phi$, that is,  at any epoch, $\gs(t)\leq1$. 
For this specific class of theories,  the lensing potential reduces to
$\Sigma=G_{\rm sc}/G_{ N}$.  
From our earlier results (see 
Figure~\ref{fig:geff_g3}, right panels), we see that 
this observable cannot be larger than unity at any cosmic epoch, and
must be  equal to 1 (the  $\Lambda$CDM value)  at the
present time.

When additional degrees of freedom are allowed (like in~\emph{H3}
and/or~\emph{H45} models),  we still find distintive features in the
evolution of $\gs$ and $\Sigma$. Indeed, the slip parameter is always smaller than unity at any redshift except possibly in the window $0.5<z<1$, where
relevant deviations from the $\Lambda$CDM expectations can be
observed. Moreover, 
$\eta$ is never larger than $\sim 1.5$ at any cosmic epoch.  
Similar to the case of \emph{BD}-like theories,  the lensing potential
$\Sigma$ is weaker than the standard model value at high redshifts
($z>0.5$--$1$), but becomes stronger (greater then unity) in recent
epochs. 
Indeed, virtually all \emph{H3} and \emph{H45} models predict an
amplitude of $\Sigma$ greater than 1 at 
the present time.

\subsection{Comparison with ad-hoc phenomenological parameterizations}

 Given the lack of a compelling model to rival GR, it is common to parameterize deviations from GR in a phenomenological manner  
 \cite{Pogosian:2010tj, simp, Ade:2015rim}.  Previous works used simple monotonic time evolution for the relevant large-scale structure observable $O(t)$ (be it $G_{\rm eff}/G_N$, $\Sigma$ or $\eta$).    
 For instance,
 the time evolution is often chosen to be proportional to the
 effective 
 dark energy density implied by the background dynamics, 
 \begin{equation}
 O(t)=1+ O_0 (1-x).
 \label{modx}  
 \end{equation}
 \noindent
so that the modified gravity scenario converges to the standard picture at high redshifts.  
Note that such a simple ansatz corresponds to straight lines in our figures~\ref{fig:geff_g3} and \ref{fig:slip}, a time dependence that does not seem to correspond to any viable single scalar field model. Instead, it appears from our analysis as if there is an additional  time scale, roughly located  between  $z=0.5$ and $z=1$. All relevant quantities undergo a transition at that scale, typically reaching a local maximum ($\eta$) or a minimum ($\geff/\gn$ and $\Sigma$). 
This behavior is not captured by the ``flat" parameterization~\eqref{modx}.

\begin{figure}[t] 
\begin{center}
\hspace{-8mm}
   \includegraphics[scale=0.75]{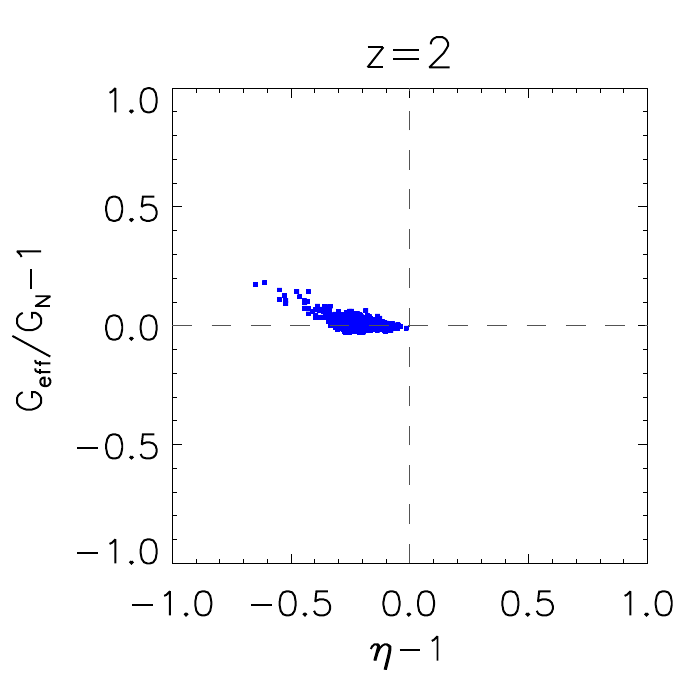}       \hskip-9mm
   \includegraphics[scale=0.75]{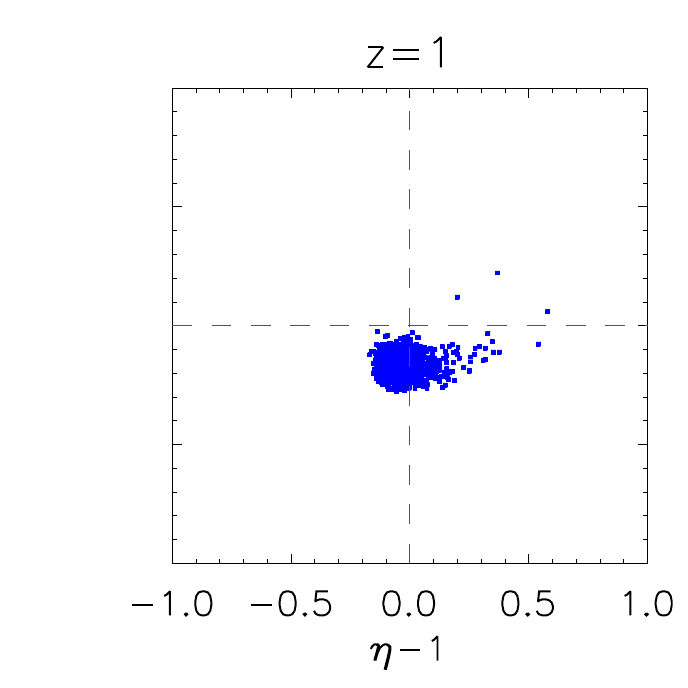}      \hskip-2mm
   \includegraphics[scale=0.70]{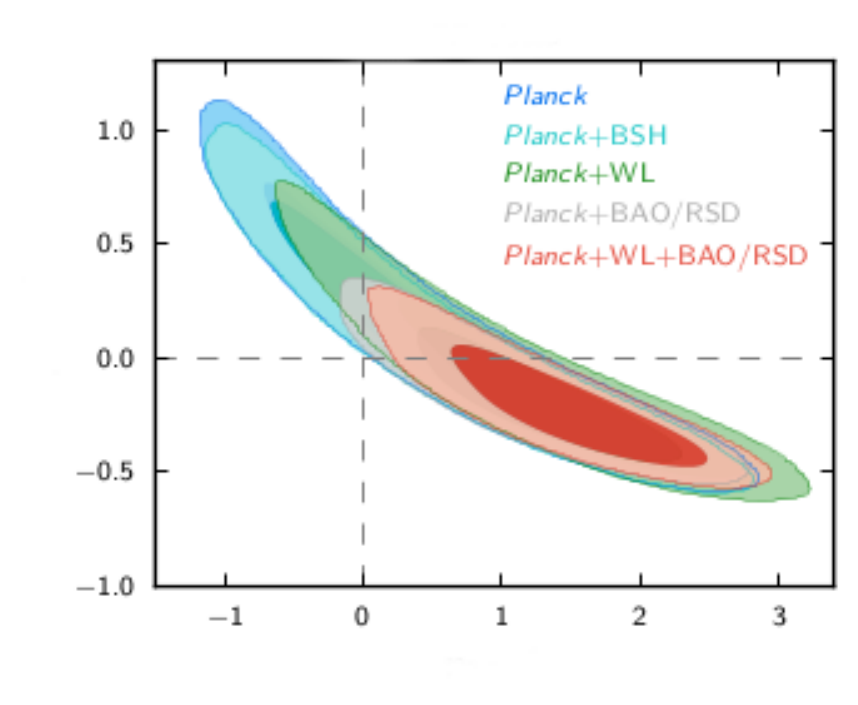}
   \hskip-67mm
   \includegraphics[scale=0.70]{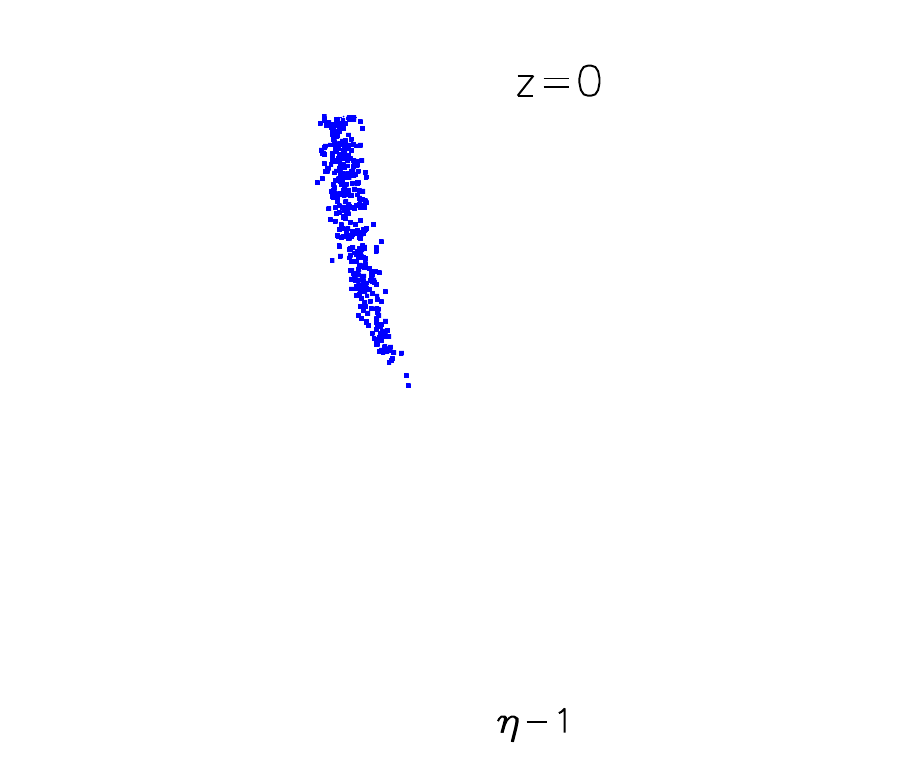}  \vskip-2mm
      \hspace{-5mm}
   \includegraphics[scale=0.75]{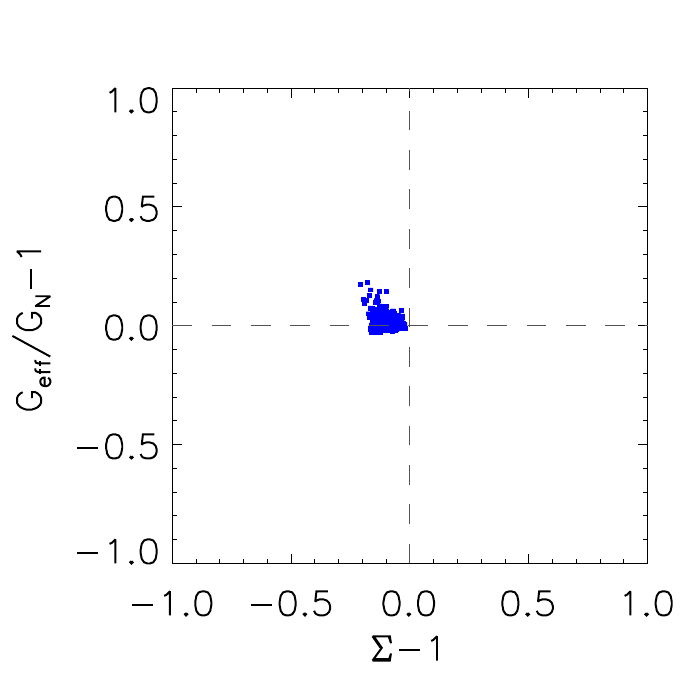}\hskip-9mm        
   \includegraphics[scale=0.75]{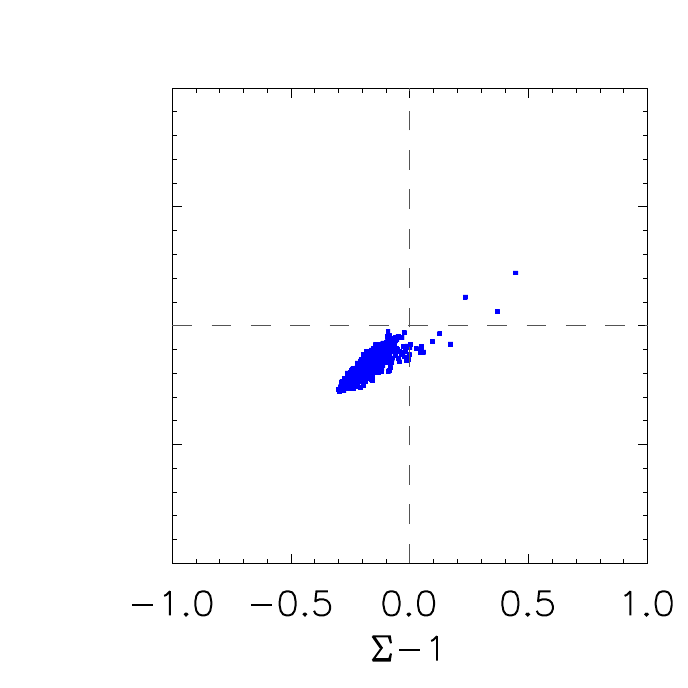}     \hskip-2mm
   \includegraphics[scale=0.70]{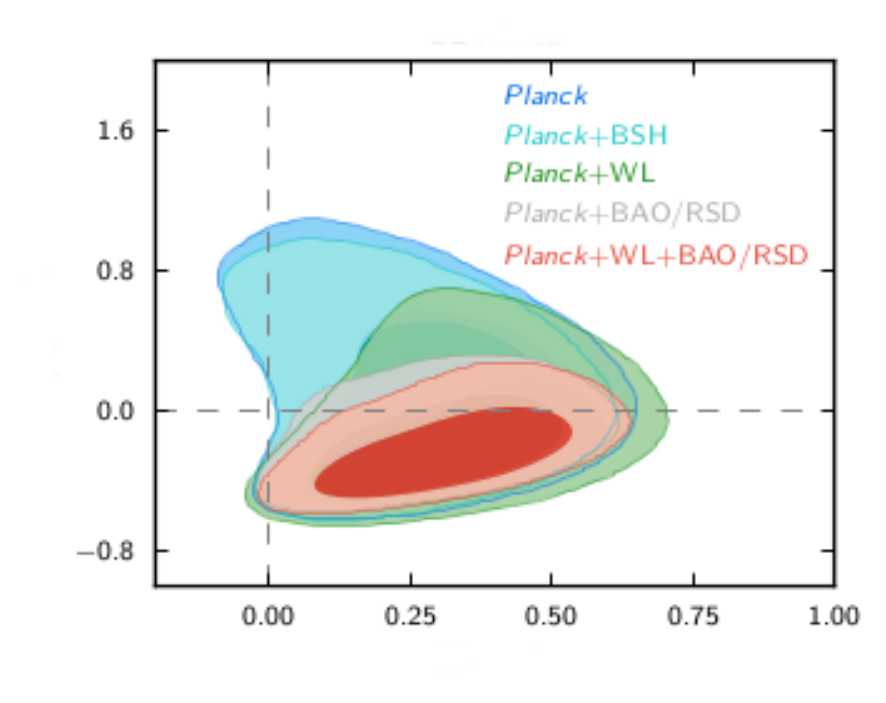}
   \hskip-59.5mm 
   \includegraphics[scale=0.70]{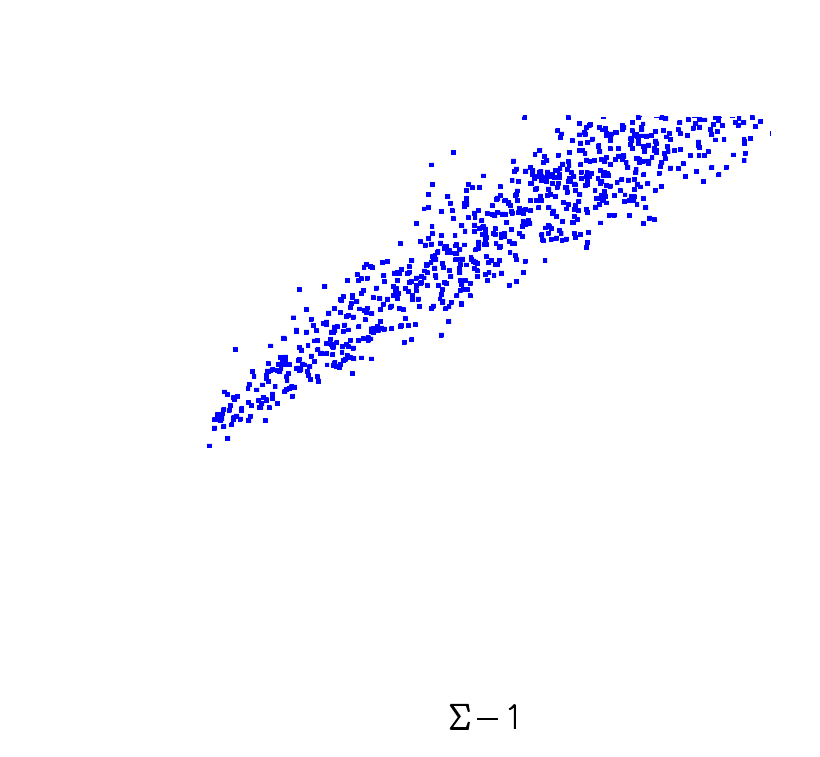}    
\caption{\label{fig:correl}  
 {\it Left and central panels}: Correlations between $\geff/\gn-1$ and $\gs-1$ (top line), $\geff/\gn-1$ and $\Sigma-1$ (bottom line) at redshifts 1 and 2 for $10^4$ stable \emph{H45} models. {\it Right panel}: The locus of points of the stable modified gravity theories encompassed by the EFT formalism is also shown at redshift equal to 0.   The 2D likelihood constraints obtained by Planck \cite{Ade:2015rim}  using a phenomenological parameterisation for the time evolution of $\Sigma$ and $\eta$ similar to that shown in eq. \eqref{modx}  is overplotted.}
\end{center}
\end{figure}

More interestingly, once the EFT formalism  is fully and consistently applied,   $G_{\rm eff}/G_{ N}$  and $\eta$---as well as $G_{\rm eff}/G_{ N}$ and $\Sigma$---appear not to be functionally independent quantities. A strong correlation exists among these pairs of observables which, in addition, 
evolves as a function of time. This correlation  is displayed in Figure \ref{fig:correl} at three different redshifts  for the general class  of \emph{H45} models.
This is yet another instructive example of the general  property
already discussed  in paper I, i.e. the  requirement of physical
stability significantly narrows the region of the parameter
space. Despite the large number of  degrees of freedom encompassed by
the EFT formalism, model predictions are tightly confined  within a
limited region in the 2-dimensional planes  $G_{\rm eff}/G_{N} - \eta$
and $G_{\rm eff}/G_{ N} - \Sigma$ at a given redshift.

Existing constraints on $\G_{\rm eff}/G_{ N}$,  $\Sigma$ and $\eta$
should be interpreted with care: they are sensitive to assumptions
made concerning their time evolution.
This is illustrated in the last panel of Figure \ref{fig:correl}. 
We superimpose the observational constraints  on these variables set by Planck
\cite{Ade:2015rim}---obtained by assuming the ansatz~\eqref{modx} for
their time dependence---with the final (at $z=0$) values of the same
quantities simulated with $10^{4}$ randomly generated viable
theories. 
Fig.~\ref{fig:correl}, interpreted at face-value, suggests viable EFT
models---with their characteristic patterns shown in  Figures
\ref{fig:geff_g3} and \ref{fig:slip}---span a region of the parameter
space that is mostly excluded by data.
Such an interpretation is unwarranted. 
Suitable (non-monotonic) time-dependence of $\G_{\rm eff}/G_{N}$,  $\Sigma$ and
$\eta$, in accordance with predictions from viable EFT models, must be
included when performing the likelihood analysis \cite{MPS}.

\section{Discussion and conclusions}
\label{Conclusions}
  
We explore the phenomenological implications of a large class of dark energy models that result
from adding a single scalar degree of freedom to the Einstein
equations with equations of motion at most second order and 
allowing for non-minimal coupling with the metric. 
To do so, we exploit the effective field theory (EFT) of dark
energy. It describes with a unifying language different,
apparently unrelated, modified gravity theories, and allows us to 
directly connect these theories to observational signatures of departure from
general relativity.
By enforcing stability and (sub-)luminality constraints, 
we ensure whatever observational signatures that emerge are
from theories free of physical pathologies.

In paper I the effective field theory of dark
energy was used to predict the linear growth index of matter density perturbations.
In this paper, we extend the formalism to compute a larger set of
perturbation observables, 
including the effective Newton constant, the linear growth rate of
density fluctuations, the lensing potential and the gravitational slip
parameter. They are expressed in terms of the structural functions of
the EFT formalism. Physical viability (i.e. freedom from pathologies) 
places strong constraints on these structural functions, greatly
enhancing the predictive power of this approach.

In particular, the time dependence of $G_{\rm eff}(t)$ (the effective
gravitational coupling in the Poisson equation \ref{pi}), $f(t)$  
(the growth function; eq. \ref{lgrowth}),
and $\eta(t)$ (the slip parameter; eq. \ref{postn}) 
for all Horndenski theories can be economically described in terms of
three constant parameters  ($x_0, w_{\rm eff}$, and $\kappa$;
eqs. \ref{xdef}, \ref{kappa}) which control the  evolution of the background
metric, as well as three functions ($\mu(t), \mu_3(t)$ and
$\epsilon_4(t)$; eq. \ref{actionn})  which, being active in the perturbation sector,  determine how matter fluctuations evolve in time at linear level.  
In this study  we are interested in modified gravity models that are kinematically equivalent, {\it i.e.} share the same expansion history,  but  
differ in their dynamical properties, {\it i.e.} predict different growth histories for large-scale structures.
We thus  factor  out the background contributions to the amplitude of the large-scale structure observables by simply fixing the three constant parameters
$x_0, w_{\rm eff}$, and $\kappa$. We do this by requiring that  the  background expansion rate  $H(t)$ be that of a $\Lambda$CDM model,  
notably the model that best fits Planck data \cite{planck}. The space of modified gravity theories is thus generated by the  three
non-minimal coupling functions $\mu(t)$, $\mu_3(t)$ and
$\epsilon_4(t)$.  To proceed further,  we Taylor 
expand each 
EFT function, and scan the theory space by generating the
expansion coefficients in a Monte Carlo fashion.
An important point is that not all the models produced in this way are viable. There is a set of stability conditions that must be satisfied in order for a 
modified gravity theory to be healthy. These act as a severe selection
on the parameter space and allow us to identify distinctive qualitative
features of single scalar field dark energy models. 
We consider $10^4$ models surviving the stability criteria while
matching the same cosmic expansion history.

This paper complements paper I in providing a maximal coverage of
the viable theory space, and strengthening the robustness and
generality of the predictions.
We confirm two central findings of paper I, i.e.  a) the space of
theories naively allowed by cosmological data is much reduced once the
viability criteria are applied;
b) the vast majority of stable Horndeski theories  produce an overall
growth of structure at 
low redshifts that is weaker than that expected in a $\Lambda$CDM
model. 
Here is a summary of our most important findings.

\begin{itemize}
\item Modified gravity does not necessarily imply \emph{stronger} gravitational attraction.
There are two effects that work in opposite directions.
One effect is best seen in the Einstein frame---the frame where the metric
is de-mixed from the scalar---the addition of an attractive
scalar force indeed makes gravity stronger. This is equivalent to the statement that
$G_{\rm eff} (t) > G_{\rm sc} (t)$
(see middle panels of Fig. \ref{fig:geff_g3}). Note that this involves a ratio of the
effective $G$ (that includes the effect of the scalar) and the
``Einstein'' $G$ (i.e. $G_{\rm sc}$) {\it at the same time}.
A separate effect is encoded in the time-evolution of the bare Planck
mass $M(t)$ (or of $G_{\rm sc}$). Models that achieve cosmic
acceleration by virtue of a genuine modified gravity effect do so by
giving $G_{\rm sc}$ a smaller value at intermediate redshifts ($z \sim
0.7$) than the current one, i.e. $G_{\rm sc}(t) < G_{\rm sc} (t_0)$.
When comparing structure growth in such models against that in
general relativity, we are really asking the question: what is
the ratio $G_{\rm eff} (t)/G_{\rm sc} (t_0)$, since $G_{\rm sc} (t_0)$
is the value of the gravitational coupling observed in the solar
system today (we call this $G_{ N}$). 
The ratio $G_{\rm eff} (t)/G_{\rm sc} (t_0)$ can
be written as $[G_{\rm eff} (t) / G_{\rm sc} (t)] \times [G_{\rm sc}
(t) / G_{\rm sc} (t_0)]$, hence it is a competition between
two factors one larger, and one smaller, than unity.
Indeed, we find that in a narrow redshift range around $z\simeq 0.7$
virtually \emph{all} MG models display $G_{\rm eff}\leq G_{ N} =
G_{\rm sc} (t_0)$. This feature is present also in the growth function
$f$, though to a lesser degree because $f$ is an integrated quantity,
dependent also on the behavior of $G_{\rm eff}$ at earlier cosmic
times. In the redshift range $0.6 \lesssim z\lesssim 0.8$, only $1\%
(/5\%)$ of modified gravity models predict a value of  $f(/f\sigma_8)$
that is larger than the $\Lambda$CDM value. 
Recent observations suggest that the growth of structure in the low
redshift universe is somewhat weaker than is expected in the best-fit
$\Lambda$CDM model. It remains to be seen if this is confirmed by
higher precision measurements in the future. It is nonetheless
interesting that a weaker growth is predicted, at intermediate
redshifts around $0.7$ or so, by the EFT of dark energy without
fine tuning of the parameters.

\item We identify an epoch of super-growth (i.e. faster growth than in
  $\Lambda$CDM) for cosmic structures. 
  At sufficiently high redshifts,  it is the first factor in
  $[G_{\rm eff} (t) / G_{\rm sc} (t)] \times [G_{\rm sc} (t) / G_{\rm
    sc} (t_0)]$    
  that wins over the second factor. At high redshifts, there is no
  need for cosmic acceleration, thus $G_{\rm sc} (t) / G_{\rm sc}
  (t_0)$ returns to unity, while $G_{\rm eff} (t) / G_{\rm sc} (t)$
  remains larger than unity due to the presence of the scalar force.
  Indeed, as already 
  demonstrated analytically in~\cite{pheno} for a more restricted number of models than is considered here, $d\geff(z)/dz$ is negative at the 
  big-bang, implying a stronger gravitational attraction at high
  redshifts. More concretely, we find that for redshifts greater than
  $z=2$, both $f$ and $f\sigma_8$ are  always larger than their
  $\Lambda$CDM values. 
  This  prediction could be used  to
  potentially rule out the whole  class of EFT models investigated in
  this paper.  Structure growth at such high redshifts is poorly
  constrained by current data, but the situation will improve with
  future surveys such as Euclid, DESI or SKA.

\item For Brans-Dicke theories, the gravitational slip parameter is at
  most one: $\gs \leqslant 1$. 
  This can be shown analytically from Eq. (\ref{postn}), or
  qualitatively understood as
  follows. Absent anisotropic stress, $\Phi = \Psi$ in Einstein frame.
  The Jordan frame $\Phi$ and $\Psi$ are obtained from their Einstein
  frame counterparts by a conformal transformation involving the scalar $\pi$.
  Because $\Phi$ and $\Psi$ comes with opposite signs in the metric,
  the conformal transformation affects them differently, leading
  to $\gs - 1\equiv (\Psi-\Phi)/\Phi \propto \pi/\Phi$, with a
  proportionality constant that makes $\gs - 1 \leqslant 0$
  (recalling that $\pi \propto \Phi$ according to the equation of
  motion; see eq. \ref{piPhi}).
We find that this feature is substantially inherited by all theories,
except for a narrow redshift range around $0.5\lesssim z\lesssim1$ when
$\gs$ exceeds one. For all theories we find  $\gs <2$ at any redshift.

\item 
The  requirements of  stability and (sub-)luminality
help greatly in narrowing the viable parameter space of dark energy theories.
Thus, despite the presence of several free functions in the EFT
formalism, the class of theories we study is highly predictive.
Moreover, the background expansion history
is already well-constrained by current data, further limiting the
form of the free functions.
Applying the EFT formalism, we find interesting correlations between
different 
large scale structure observables.
In particular, at a given redshift, there exists only a narrow region 
in the 2-dimensional planes  $G_{\rm eff}/G_{ N} - \eta$ and $G_{\rm
  eff}/G_{ N} - \Sigma$ where data can be meaningfully interpreted in
terms of  viable theories. 
\end{itemize}


The systematic investigation by means of the EFT formalism of what
lies beyond the standard gravity landscape is still in its infancy, and a number of 
improvements would be desirable.
For example, it would be interesting to 
work out the consequences of relaxing some of the constraints imposed
in our current analysis. How might our predictions be modified
if a different initial condition were chosen ($\kappa \ne 1$), or
if the background expansion history is altered ($w_{\rm eff} \ne
-1$)? It would also be interesting to consider early dark energy models, in which a relevant fraction of dark energy density persists at early times (\emph{i.e.} not imposing eqs.~\ref{limitM} and~\ref{constrain}). 
We have focused on scales much smaller than the Hubble radius
in this paper. As data improve on ever larger scales, our analysis
should be extended to include possible scale dependent effects 
that come from mass terms for $\pi$ that are of the order of Hubble.
Lastly,  it would be useful to isolate models that have genuine
self-acceleration, and compare their predictions with the ones
studied here (which include both models that self-accelerate and
models that do not).

\section*{Acknowledgments}
We acknowledge useful discussions with Julien Bel, Emilio Bellini, Jose Beltran Jimenez, Kazuya Koyama,
Valeria Pettorino, Valentina Salvatelli, Iggy Sawicki, Alessandra
Silvestri, Heinrich Steingerwald, Shinji Tsujikawa, Hermano Velten, Filippo Vernizzi and Jean-Marc Virey.
F.P. warmly acknowledges the financial support of A*MIDEX project
(n$^{o}$ ANR-11-IDEX-0001-02) funded by the ``Investissements
d'Avenir" French Government program, managed by the French National
Research Agency (ANR). CM is grateful for support from specific
project funding of the  Labex OCEVU. LH thanks the Department
of Energy (DE-FG02-92-ER40699) and NASA (NNX10AN14G, NNH14ZDA001N) 
for support, and Henry Tye at the HKUST Institute for
Advanced Study for hospitality.

\appendix
\section{EFT action and relation with other equivalent parameterizations}\label{app-a}

While referring to the existing literature~\cite{EFTOr,Bloomfield:2012ff,GLPV,Bloomfield:2013efa,PV} for more details, here we just display the EFT action written in unitary gauge, which implicitly \emph{defines} the couplings~\eqref{couplings}, 
\be
\begin{split}
S \ = \ & \  S_m[g_{\mu \nu},\psi_i] \ \ + \int \! d^4x \sqrt{-g} \, \frac{M^2(t)}{2} \, \left[R \, -\,  2 \lambda(t) \, - \, 2 \cb(t) g^{00} \, \right. \\
 &  \left.-\,\mu_2^2(t) (\delta g^{00})^2 \, -\, \mu_3(t) \, \delta K \delta g^{00} + \,  \epsilon_4(t) \left(\delta K^\mu_{ \nu} \, \delta K^\nu_{ \mu} -  \delta K^2  +  \frac{\!\!\R\,   \delta g^{00}}{2}\right) \right] \;,
\end{split}
\label{action}
\ee
Here the metric is the Jordan one, to which matter fields $\psi_i$ minimally couple (we are assuming the validity of the weak equivalence principle here, see~\cite{Gleyzes:2015pma} for a relaxation of this hypothesis).
 The functions $\lambda$ and ${\cal C}$ are constrained by the Friedman equations (see Sec.~\ref{background} or Ref.~\cite{pheno}). 

In order to arrive at the action~\eqref{actionn} from~\eqref{action}
we need to change gauge. The Stueckelberg procedure (see
\emph{e.g.}~\cite{PV} for details) allows one to exit the unitary
gauge by performing a time diffeomorphism $t \rightarrow t+\pi$, where
$\pi$ represents the fluctuations of the scalar field, that now can
appear explicitly in the action and in the equations. The Newtonian
gauge then follows from further requiring a metric of the
form~\eqref{newtonian}. Note that the Stueckelberg procedure does not
create direct coupling between $\pi$ and the matter fields, consistent
with the fact that we are working in the Jordan frame, and the matter action is diff-invariant.

Finally, it is useful to write a dictionary relating our couplings to the equivalent ``$\alpha$ couplings" defined in Ref.~\cite{Sawicki} (see also~\cite{Gleyzes:2014rba}):
\be
\begin{split}
M^2_* & =M^2\, (1+ \epsilon_4)\, , \\
\alpha_M & =\frac{\dot \epsilon_4}{H(1+ \epsilon_4)} + \frac{\mu}{H}\, , \\
\alpha_K & = \frac{2{\cal C}+4\mu_2^2}{H^2 (1+ \epsilon_4)}\, , \\  
\alpha_B & =\frac{\mu_3-\mu}{H (1+ \epsilon_4)}\, , \\
\alpha_T  & =-\frac{\epsilon_4}{1+\epsilon_4} \;. 
\end{split}
\ee

\section{Exploring EFT models by means of  a  different parameterization scheme}\label{wparam}

In paper I a specific parametrization of the Brans-Dicke non-minimal coupling $\mu(t)$ was suggested, based on the  ``physical" equation of state parameter for dark energy $w \, = \, p_{D}/\rho_{D} \,$. This approach provides a complementary  way to solve the background/perturbation sector degeneracy in the action \eqref{actionn} so that the function $\cb$ is no longer free but is directly linked to $\mu$.  The details about  this parameterization scheme (the so called ``$w$-parameterization" in opposition to the ``$\mu$-parameterization"  adopted in the current paper) can be found in Paper I.  We 
briefly summarize here  only the intermediate steps that are needed to relate these two parameterizations, and are thus important for expressing 
observable such as  $\geff$ and $\fs$ in the language of paper I.

A little bit of algebra is enough to show that  $\mu$ and  the {\it physical} equation of state parameter $w$ are related as follows
\begin{equation}\label{muw}
 \mu(x)=\dfrac{3\wb H(1-x)}{w-\wb(1-x)}\left[ w-\wb+ x(1-x)\frac{\wb}{w}\dfrac{dw}{dx}\right]\, .
\end{equation}

\begin{figure}[H]
\centering \hskip-2mm
   \includegraphics[scale=0.76]{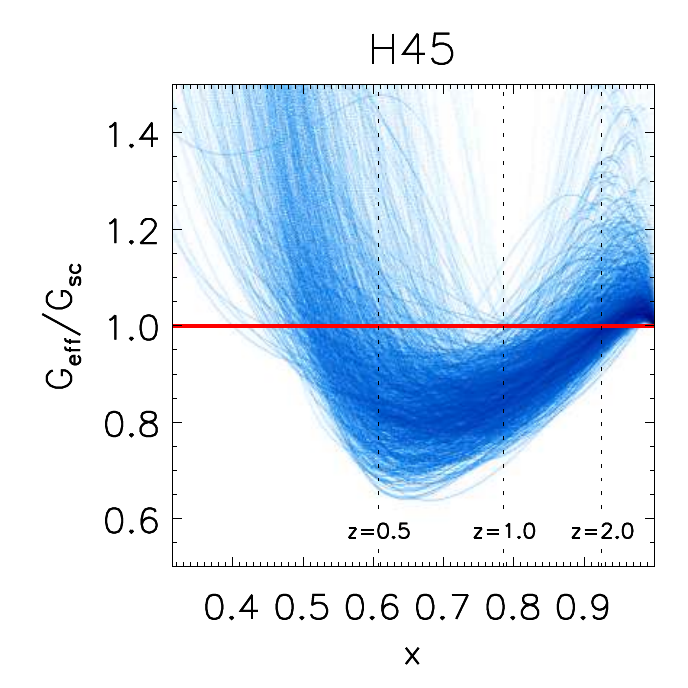}  \hskip-3mm
   \includegraphics[scale=0.76]{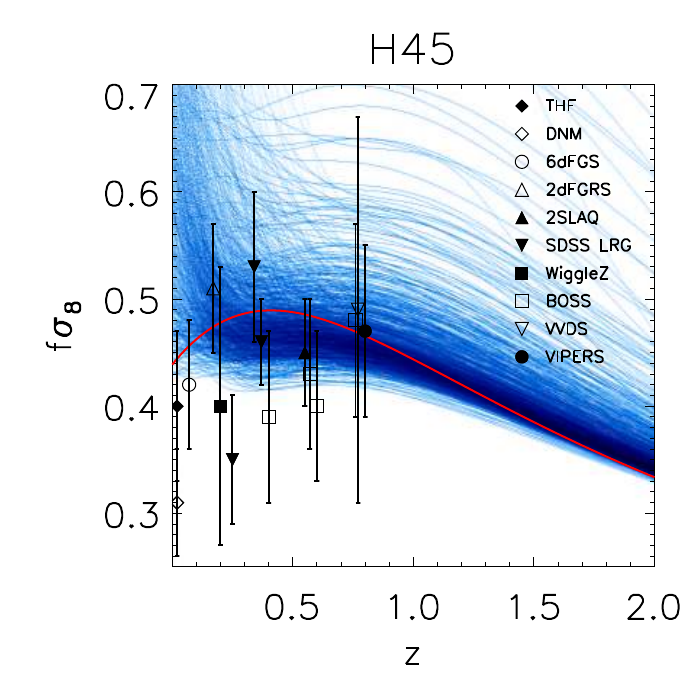}   \hskip-1mm
   \includegraphics[scale=0.76]{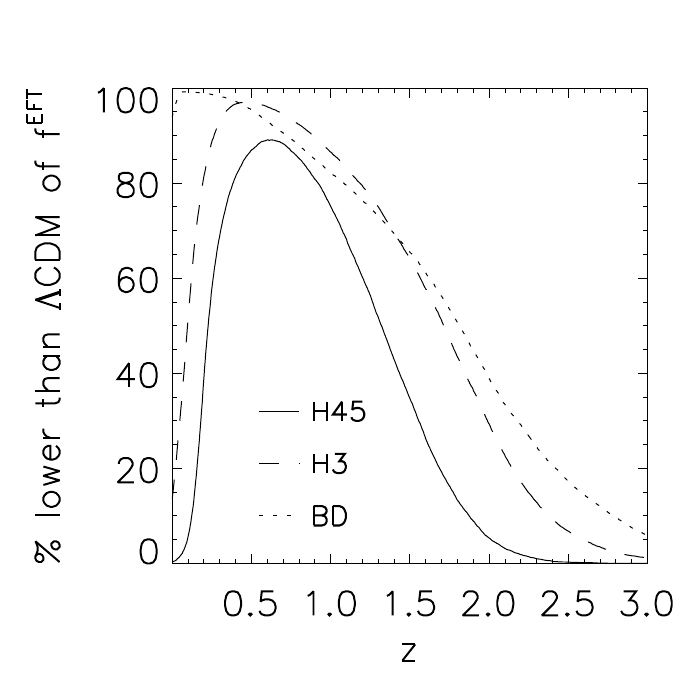} 
\caption{The plots displayed are computed in the $w$-parameterization
  scheme. The evolution of $\geff(x)$ (left), $\fs(z)$ (middle) for a
  sample of $10^3$ randomly generated models of viable \emph{H45}
  theories is shown. The percentage of $10^4$ randomly generated EFT models with growth rate $f$ \emph{lower} than that predicted by $\Lambda$CDM Planck-cosmology as a function of the redshift $z$ (right) is also shown. The thick red line represents the prediction of $\Lambda$CDM.} 
   \label{fig:plots_wparam}  
\end{figure}

One can  thus proceed by expanding in Taylor series the function $w$ (instead of directly $\mu$, as we  have done in this paper). 
\begin{align} \label{wtay1}
w\left(x\right)\ &=  \ \wb\,\dfrac{1-x_0}{1-\kappa\,x_0\,(1+ \epsilon_4^0)^2}+\pu_1 \left(x -x_0\right)+\pd_1 \left(x -x_0\right)^2 \ ,
\end{align}
We should mention that the zero-th order term is fixed by requiring that the expansion rate of the various EFT models be identical to that 
of the reference  flat $\Lambda$CDM scenario ( $\wb=-1$ and $x_0=0.314$).   
The other couplings functions are expanded as discussed in section 3.1 of the present paper.  
\vv
Figure \ref{fig:plots_wparam} displays some of the results we obtain by implementing  the $w$-parameterization scheme. Since there is no natural unit for the coupling functions, we chose to randomly pick $p_1^{(i)} \in [-0.8,0.8]$ (for $w$) and $p_{2,3}^{(i)} \in [-2,2]$ (for $\mu_3$ and $\epsilon_4$) in order for $\fs$ curves to cover current observational  data as much as in the $\mu$-parameterization. The resemblance of $\geff/\gn$ to the one presented in Figure \ref{fig:geff_g3} confirms the robustness of our general findings; the universal behavior of $\geff$ ({\it S-shape}) is recovered no matter what parameterization is used for the coupling functions. In addition, the fraction of models with lower/higher growth with respect to $\Lambda$CDM displays the same general scaling as a function of $z$. We find that  $\sim 90\%$ of \emph{H45} theories predict growth rates below the $\Lambda$CDM prediction  in the redshift window  $0.5\lesssim z \lesssim 1$. The change of parameterization implies a slight modification of the asymptotic behavior of the coupling functions $\mu$ and $\cb/H^2$ at $x=1$, more precisely a modification of the speed at which these two functions go to zero during mater domination. This is well illustrated by the tail of the curves, the transition to the super-growth epoch is now shifted to a somewhat  higher redshift ($z\sim 2.5$).

\end{document}